\documentclass[superscriptaddress,hyperref,showpacs,showkeys]{revtex4-1}

\usepackage{graphicx}
\usepackage{amsmath}
\usepackage{amsfonts}
\usepackage{amssymb}

\makeatletter
\def\p@subsection{}
\def\p@subsubsection{}
\makeatother

\newcommand{\fref}[1]{Fig. \ref{#1}}
\newcommand{\sref}[1]{Sec. \ref{#1}}
\newcommand{\eref}[1]{(\ref{#1})}

\newcommand{\avg}[1]{\left\langle #1 \right\rangle} 

\begin{document}
\title{BITLLES: Electron Transport Simulation with Quantum Trajectories}

\author{Guillermo Albareda}
\email[]{albareda@ub.edu}
\affiliation{Materials Science and Physical Chemistry \& Institute of Theoretical and Computational Chemistry, Universitat de Barcelona, 08028 Barcelona, Spain}

\author{Damiano Marian}
\affiliation{Depertament d'Enginyeria Electr\`onica, Universitat Aut\`onoma de Barcelona, Bellaterra, Spain}
\author{Abdelilah Benali}
\affiliation{Depertament d'Enginyeria Electr\`onica, Universitat Aut\`onoma de Barcelona, Bellaterra, Spain}
\author{Alfonso Alarc\'on}
\affiliation{Depertament d'Enginyeria Electr\`onica, Universitat Aut\`onoma de Barcelona, Bellaterra, Spain}
\author{Simeon Moises}
\affiliation{Depertament d'Enginyeria Electr\`onica, Universitat Aut\`onoma de Barcelona, Bellaterra, Spain}
\author{Xavier Oriols}
\email[]{xavier.oriols@uab.cat}
\affiliation{Depertament d'Enginyeria Electr\`onica, Universitat Aut\`onoma de Barcelona, Bellaterra, Spain}



\begin{abstract}
 After the seminal work of R. Landauer in 1957 relating the electrical resistance of a conductor to its scattering properties, much progress has been made in our ability 
 to predict the performance of electron devices in the DC (stationary) regime. 
 Computational tools to describe their dynamical behavior (including the AC, transient and noise performance), however, are far from being as trustworthy as would be desired 
 by the electronic industry. 
 While there is no fundamental limitation to correctly modeling the high-frequency quantum transport and its fluctuations, certainly more careful attention 
 must be paid to delicate issues such as overall charge neutrality, total current conservation, or the back action of the measuring apparatus.  
 In this review, we will show how the core ideas behind the Bohmian formulation of quantum mechanics can be exploited to design an efficient 
 Monte Carlo algorithm that provides a quantitative description of electron transport in open quantum systems. By making the most of trajectory-based and wave function methods, 
 the BITLLES simulator, a free software developed by the authors, extends the capabilities that the semi-classical Monte Carlo simulation 
 technique has offered for decades (DC, AC, noise, transients) to the quantum regime. 
\end{abstract}

\maketitle
\tableofcontents

\section{Introduction: Why Bohmian Mechanics?}
\label{Intro}

There are many different formulations of classical mechanics (Newtonian, Lagrangian, Hamiltonian, Poisson Brackets, Hamilton-Jacobi, etc.). The various formulations differ mathematically and conceptually, yet each one makes identical predictions for all experimental results. The situation in quantum mechanics is certainly very similar.  For example, elementary textbooks teach us that the harmonic oscillator problem is cleanly and easily solved through the creation and annihilation operators of the matrix (Heisenberg) formulation, while many other simple problems are better formulated directly with the wave function (or Schr\"{o}dinger equation) \cite{o.cohen1978book}. Another relevant example is the (Feynman) path integral formulation which is rarely the easiest way to approach a non-relativistic quantum problem, but it has innumerable and very successful applications in quantum field theory \cite{o.feynmann1965book}.  Some problems look difficult in one formulation (interpretation) of quantum mechanics and easy in another. The aim of this chapter is to discuss in which extent one of the formulations (interpretations) of the quantum theory, i.e. Bohmian mechanics, provides a singular tool to predict and explain the behaviour of novel electronic devices (an extended revision of the use of Bohmian mechanics to solve practical problems can be found in \cite{AppBohm}).

Bohmian mechanics was originally proposed by Louis de Broglie in 1924 \cite{o.debroglie1923} and fully developed into a consistent explanation of all quantum phenomena in terms of wave and particles by David Bohm in 1952  \cite{o.Bohm1952a}. There are many good references where its basic ingredients can be easily understood \cite{o.bell2004book,o.bohm1993book,o.durr2012book,o.durr2009book,o.oriols2011book}. Nevertheless, since its mathematical structure for the non-relativistic quantum mechanics used in this chapter is quite simple, the brief introduction in \sref{Bohm} is enough to understand the whole chapter (a connection with other
simulation tools mentioned in this book can be found in Appendix D where a formal relation between the Wigner distribution function and the Bohmian trajectories is established).  

Bohmian mechanics agrees with all quantum experiments done up to now. All researchers who analyze the ideas of de Broglie and Bohm with the pertinent scientific rigor conclude that there is no objective argument against them\footnote{John S. Bell used the following words to explain how simply the iconic double-slit experiment, in particular, and any other quantum phenomenon, in general, can be understood with Bohmian mechanics: ``While the founding fathers agonized over the question \emph{particle} or \emph{wave}, de Broglie in 1925 proposed the obvious answer 'particle' and 'wave'. Is it not clear from the smallness of the scintillation on the screen that we have to do with a particle? And is it not clear, from the diffraction and interference patterns, that the motion of the particle is directed by a wave? De Broglie showed in detail how the motion of a particle, passing through just one of two holes in screen, could be influenced by waves propagating through both holes. And so influenced that the particle does not go where the waves cancel out, but is attracted to where they cooperate. This idea seems to me so natural and simple, to resolve the wave-particle dilemma in such a clear and ordinary way, that it is a great mystery to me that it was so generally ignored''. \cite{o.bell2004book}.}.
Then, \emph{why is it generally ignored by the scientific community?} There are several historical and sociological reasons  that justify its marginal status. We mention two of them:
First, there is a \emph{vicious circle} with negative feedback. Since few people knows Bohmian mechanics, few people uses it for practical applications. Then, there are no much Bohmian contributions and, consequently,  
this formulation is not taught at Universities or explained in research courses. As a result, few people knows it and the \emph{circle} starts again. Let us mention that the quantum chemistry community is an encouraging 
exception that has been able to escape from this \emph{vicious circle} fifteen years ago, mainly due to the works of R. Wyatt and coworkers \cite{o.wyatt2005book,o.wyatt2000jcpa,o.wyatt2000jcpb}.
Second, there is a widely spread belief that Bohmian mechanics, by construction, has a limited usefulness. It is argued that, apart from computing the wave function, Bohmian mechanics requires tracking a set of 
trajectories that, at the end of the day, will exactly reproduce the time-evolution of the wave function, which was already known. Then, \emph{What is the utility of the extra effort for computing Bohmian trajectories?} 
This criticism is valid for a single-particle problem where the wave function can be explicitly computed, but it is not pertinent at all for realistic quantum problems, where the wave function itself cannot 
be computed (because it lives in a $R^{3N}$ configuration space). The so-called many-body problem\footnote{P.A.M. Dirac, wrote in 1929: ``The general theory of quantum mechanics is now almost complete. 
The underlying physical laws necessary for the mathematical theory of a large part of physics and the whole of chemistry are thus completely known, and the difficulty is only that the exact application of these 
laws leads to equations much too complicated to be soluble \cite{o.dirac1929}.''}. In fact, any practical formulation of quantum transport is built up independently of the many-particle wave function in the configuration space and 
it evolves around some mathematical entity living in the real space. For Bohmian mechanics, the conditional (Bohmian) wave function, described in \sref{Bohm3}, is such entity. 
Apart from the ability to approximate many-body problems with conditional wave functions defined in the real space, Bohmian mechanics allows a simple description of the quantum measurement process without invoking any special postulate 
for the collapse.

Because of the previously mentioned advantages, we argue that Bohmian mechanics is a well-suited computational tool for studying quantum transport,  in general, and its high-frequency behaviour, in particular. 
This effort on accurately predicting high frequency quantum transport is somehow urgent nowadays because the International Technology Roadmap for Semiconductors (ITRS\footnote{{http://www.itrs.net} }) 
is expecting that quantum devices with THz operating frequencies will play an important role in the future electronic industry within a few years.  
For those readers familiar with electron transport with the Monte Carlo solution of the Boltzmann Eq. \cite{o.jacoboni1989book}, Bohmian trajectories play the same role of the semi-classical Monte Carlo trajectories, 
but in a rigorous quantum regime. They provide a microscopic description (in terms of well-defined trajectories guided by waves) of the ensemble results obtained from other (wave alone) formulations.
The rest of this introduction is devoted to revisit what, in our opinion, constitute two underlying difficulties that one has to face when modeling electronic devices beyond the DC regime. 
Such difficulties are the quantum measurement and the many-body problems. Along the whole chapter we will describe how Bohmian trajectories can help to solve them. 
After this introduction, in \sref{Bohm} we summarize postulates, equations and the most important features of Bohmian mechanics regarding the description of electron transport including the measurement process and many-body correlations. 
\sref{Bitlles} describes how Bohmian mechanics can be applied to simulate electron transport. We will discuss the main pieces required to build up a trajectory-based electronic device simulator. 
Although this simulator is able to provide any kind of dynamic property related with a quantum device, we will focus on the electrical current. 
The rest of the chapter, \sref{Computation}, is expressly devoted to provide a practical method to evaluate the electrical current. 
Thus, after a brief dissertation on the computation of DC currents in quantum systems and on the role of the multi-time measurement to predict its fluctuations, we provide a detailed description of the equations 
required to evaluate, DC, AC, transients and any moment of the current. We conclude in \sref{conclusions}.

\subsection{Quantum continuous measurement of the electrical current}
\label{Intro1}

Before going into the  Bohmian formulation of quantum transport, let us start by explaining why the quantum measurement at successive times plays a crucial role in the high-frequency behaviour of quantum devices. 
This discussion will help us in understanding the practical and conceptual difficulties of high-frequency quantum transport modeling.

In 1926, E. Schr\"{o}dinger published \emph{\rq\rq{}An Undulatory Theory of the Mechanics of Atoms and Molecules\rq\rq{}} \cite{o.Schrodinger}, where he described electrons in terms of a wave function 
$\psi(\vec r,t)= \langle \vec r|\psi(t) \rangle $  solution of his famous equation. 
He interpreted the wave function  $\psi(\vec r,t)$ as a description of the electron charge density defined as $q|\psi(\vec r,t)|^2$ being $q$ the electron charge. 
Later, M. Born refined the interpretation of Schr\"{o}dinger and defined  $|\psi(\vec r,t)|^2$  as the probability density of finding the electron in a particular position $\vec r$  and time  $t$ \cite{o.waerden}.

\begin{figure}
\centerline{
\includegraphics[width=0.77\columnwidth]{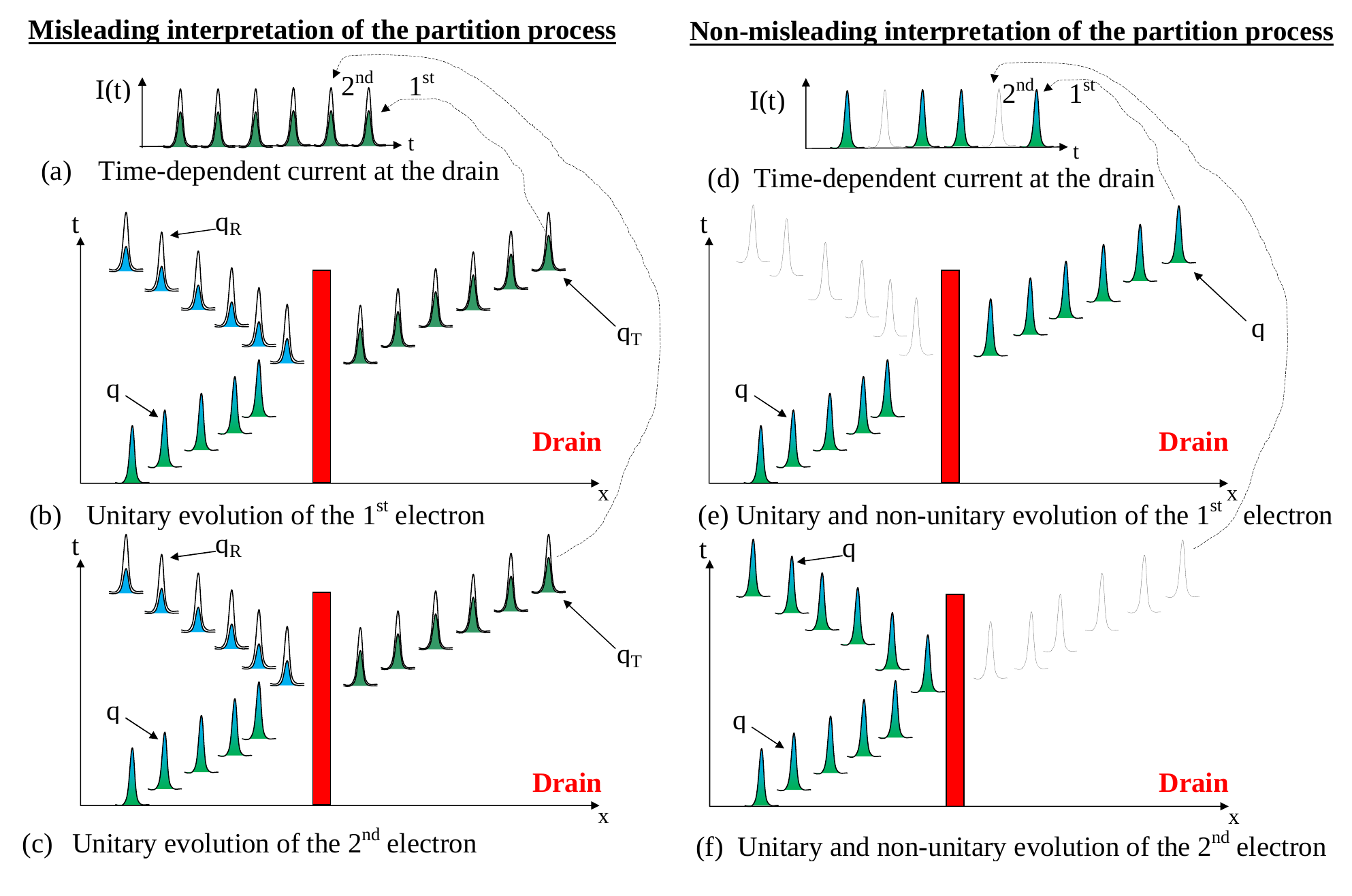}}
\caption{ (Color online) Schematic representation of the partition noise present on the measured current in (a) generated by a flux of electrons impinging on a tunneling barrier. Only the unitary (Schr\"{o}dinger-like) evolution of the wave function is considered in (b) and (c). This unitary evolution alone provides a misleading explanation of quantum transport. The correct explanation with unitary and non-unitary (collapse-like) evolution is shown in (d), (e) and (f). }
\label{o.figure1}
\end{figure}

In order to realize why the interpretation of the wave function as a charge distribution is misleading\footnote{The error of interpreting $|\psi(\vec r,t)|^2$  as a charge density is clearly seen when dealing, 
for example, with 3 electrons whose (many-particle) wave function ``lives'' in a configuration space of 9 variables, while the charge density is still defined in the real space with 3 variables. 
In scenarios with a very large number of particles, $n(\vec r,t)=N \int d\vec r_2 d\vec r_3\dots d\vec r_N|\psi(\vec r,\vec r_2,...,\vec r_N,t)|^2$  can be quite similar to the charge density, but not identical.}, 
it will be relevant to discuss the DC and the partition noise generated by a flux of electrons impinging upon a tunneling barrier. The transport process is schematically depicted in Fig. \ref{o.figure1}. 
For simplicity, we assume only injection of electrons from left to right. After the interaction with the barrier, the solution of the (time-dependent) Schr\"{o}dinger Eq. \cite{o.cohen1978book} provides a wave 
function that is spatially separated into transmitted, $\psi_T(\vec r,t)$,  and reflected, $\psi_R(\vec r,t)$, parts with $\psi(\vec r,t)=\psi_T(\vec r,t)+\psi_R(\vec r,t)$.

Naively, one could think that the charge $q$ of each injected electron is divided into two smaller parts, $q_T=q \int |\psi_T(\vec r,t)|^2 dr$  for the transmitted part and also $q_R=q \int |\psi_R(\vec r,t)|^2dr$  for the reflected one as seen in Fig. \ref{o.figure1}(a)-(c). This wrong interpretation of the wave function would produce a DC value equal to $I_{DC}=\nu \cdot q_T$, $\nu$ being the number of electrons injected per unit of time. However, a measured electron is either fully transmitted or fully reflected as seen in Fig. \ref{o.figure1}(d)-(f). Not both\footnote{The \emph{ontological} meaning of the dotted \emph{empty} waves drawn in Fig. \ref{o.figure1} depends on which interpretation of quantum mechanics is selected (we mentioned some of them in Refs. \cite{o.zurek2005,o.cohen1978book,o.Schlosshauer,o.Griffiths,o.Bohm1952a}). In any case, such selection is not at all relevant in our argument. We only want to emphasize the experimental fact that, in the ammeter, we only detect a transmitted or reflected electron, but not both. }. From the probabilistic interpretation of the wave function, the transmission coefficient $T=\int |\psi_T(\vec r,t)|^2dr$ has to be defined as the ratio of transmitted electrons over the injected ones. An identical interpretation is required for the reflection coefficient $R=\int |\psi_R(\vec r,t)|^2dr$. This correct interpretation of the wave function provides a (ensemble-average) value of the DC current equal to $I_{DC}=\nu \cdot  q \cdot  T$, which becomes equal to the previous wrong result $I_{DC}=\nu \cdot  q_T$ because $q_T=q \cdot T$.

A more formal explanation about why Figs. \ref{o.figure1}(a)-(c) are intrinsically misleading to understand quantum transport is because, in principle, one cannot use the Schr\"{o}dinger equation alone to discuss the time evolution of a quantum system. Orthodox textbooks \cite{o.cohen1978book} point out that there are two dynamical laws for quantum systems. Between two measurements, the quantum system follows a unitary time-evolution determined by the Schr\"{o}dinger equation. On the contrary, during the measurement process, the quantum system suffers a non-unitary evolution. For example, by being projected onto the measurement eigenstates. The unitary evolution in Fig. \ref{o.figure1}(a)-(c) preserves the norm of the wave function, while the non-unitary evolution in Fig. \ref{o.figure1}(d)-(f) \lq\lq{}eliminates\rq\rq{} those parts of the wave function that do not correspond to the measured data. Whenever the measurement of the current is relevant in the evolution of the quantum device, both laws are mandatory. However, for studying DC, a single measurement is enough and, then, the formal discussion on the role of measurement can be somehow relaxed (see \sref{preliminar} for a detailed discussion on this point). To the contrary, the transport process described in Fig. \ref{o.figure1}(d)-(f), which schematically includes both dynamical laws, becomes fully pertinent to understand the fluctuations of the current around its DC value. The first erroneous interpretation of quantum transport in Fig. \ref{o.figure1}(a)-(c) provides no noise (at zero frequency). At each time-interval $1/\nu$, the transmitted charge is always equal to $q_T=qT$. There are no fluctuations.  See Fig. \ref{o.figure1}(a). The second interpretation in Fig. \ref{o.figure1}(d)-(f) provides the experimental partition noise measured in the laboratory. The transmitted charge during the time-interval  $1/\nu$ fluctuates between the value $q$ , when the electron is fully transmitted with probability $T$, and the value $0$ when it is fully reflected with probability $R=1-T$. See \fref{o.figure1}(d).

The situation can be even worst if we are interested in the power spectral density. Then, as it will be discussed in \sref{recipe}, we have to consider multi-time measurements (at least, time-correlations between two values of the current measured at two different times). \emph{How do we have to model the evolution of the system during the measurement process?} The answer is certainly not simple because of the inherent difficulties of the quantum measurement process (even at an ontological level \cite{o.Schlosshauer}). In principle, the wave function has to suffer a non-unitary evolution in order to eliminate (at least modify) parts of the wave function during the measurement process\footnote{It is noticeable that a non-unitary evolution of the wave function is also useful to model the process of \emph{decoherence} in quantum systems \cite{o.Schlosshauer}. For macroscopic systems, we cannot hope to keep track of all microscopic degrees of freedom. We trace out many variables (by integrating them) in order to keep only the most relevant ones (those belonging to electrons at the active region). Then, the equation of motion of this \emph{open} system is not determined by the Schr{\"{o}}dinger equation, but by different ones that allow simultaneous unitary and non-unitary evolutions of the open system.}. As it will be shown in \sref{Bohm2} for generic measurements and in \sref{Bitlles1} to compute the electrical current in particular, Bohmian mechanics provides a straightforward solution to the measurement problem.

\subsection{Coulomb correlations and displacement current}
\label{Intro2}

The second issue that we want to discuss to emphasize the difficulties to correctly model high frequency quantum transport is the relevance of the displacement current  in such scenarios. For DC transport, we generally deal only  with the conduction current (or particle current) $I_p(t)$ related to the number of electrons crossing a particular surface $S_i$.  Nevertheless, the electric field  inside a quantum device is both inhomogeneous and time varying because of its time dependence on the external bias and the movement of electrons. Under such time-dependent scenarios, a displacement current $I_d(t)$, proportional to the time-derivative of the electric field,  is always present.  

The displacement current has no role when modeling DC because, by definition, the time-average value of $I_d(t)$ is zero. However,  $I_d(t)$ has a fundamental role when modeling high-frequency transport. The total current is $I(t)=I_p(t)+I_d(t)$. The total current has to satisfy the current conservation law, meaning that $I(t)$  evaluated on a closed surface $S$ must be zero at any time. This is just a consequence of the Maxwell equations \cite{o.jackson1962book}. This property allows us to argue that the current $I(t)$  measured by an ammeter (far from the simulation box) is equal to the current that we compute on a particular surface, $S_i$, of the simulation box. Therefore, special care must be taken with the computation of the electric field to obtain the displacement component, because the total time-dependent current cannot be known from merely counting transmitted electrons.

For electron devices, an additional clarification about the displacement current is relevant. Using the Gauss\rq{} equation, the current conservation law can be rewritten as a continuity equation $\partial \rho/\partial t+\vec \nabla \vec J_p=0$  , being  $\vec J_p$ the conduction current density and  $\rho$ the charge density. Then, if we integrate this continuity equation over a very large volume $V$ with boundaries deep inside the reservoirs, we can impose an additional charge neutrality requirement, $\int_V \rho dr^3=0$. This extra requirement is just a consequence of the fact that positive or negative deviations from charge neutrality inside the device tend to zero after time intervals proportional to the dielectric relaxation time \cite{o.albareda2010prb,JCEL_BCs}. 

Therefore, in most scenarios, the proper modeling of electron transport has to take into account the time-dependence of the external bias together with the dynamics of electrons in a self-consistent way. This requires an (approximate) solution to the quantum many-body problem. The origin of this many-body problem is that, in the quantum world, the wave function that determines electron dynamics is defined in the N-particle configuration space, $\Psi(\vec r_1,\vec r_2,...,\vec r_N,t)$. Therefore, the exact treatment of the Coulomb interaction among electrons in principle has to be defined in the N-particle configuration space as well, with a potential energy of the type $U(\vec r_1,\vec r_2,...,\vec r_N,t)$. Solving the problem in the N-particle configuration space represents a very difficult (impossible) computational task. As a result, only reasonable approximations for the many-body problem are computationally accessible\footnote{The computational burden becomes even more dramatic if we include not only (free) electrons in the transport model but also the atoms that define the electronic band structure. Indeed, as devices are approaching the few-atoms limit, the quantum transport modeling becomes more and more interwoven with material modeling.}. We will see in \sref{Bohm3} that the Bohmian formulation in terms of conditional wave functions provides an approximate solution to the many-body problem.

\section{Theoretical Framework: Bohmian Mechanics}
\label{Bohm}

Bohmian mechanics is, most of the times, introduced through an alternative set of (non-linear) equations that, together, play the role of the (linear) Schr\"odinger one. First, the continuity equation for the probability distribution leads to the
definition of the Bohmian velocity field. Second, a modified Hamilton-Jacobi equation is found after introducing the so called \emph{quantum potential} \cite{o.bell2004book,o.durr2012book,o.oriols2011book}. 
In this work, however, we will directly deal with the Schr\"odinger equation and its conditional form \cite{o.oriols2007prl,PRL_Alb, JPCL}.

In order to make the reader aware of the Bohmian formalism, in \sref{Bohm1} we first introduce in a very compact way the basic postulates and equations useful for this work. In \sref{Bohm2} we explain why and in which way Bohmian mechanics constitutes a quantum theory without observers, i.e. the quantum measurement in the Bohmian theory is described without invoking the wave function collapse postulate. Finally, in \sref{Bohm3} we explain in which way the concept of \emph{conditional wave function} can be useful to tackle the many-body problem.

\subsection{Postulates and basic equations}
\label{Bohm1}

Consider a system of $M_T$ (spinless) electrons described in the spatial coordinates $\vec r = \{\vec r_1,...,\vec r_{M_T}\}$ by the many-particle wave function $\Psi(\vec r,t)$ which obeys the many-particle Schr\"odinger equation, i.e.:
\begin{equation} \label{Scho}
i\hbar \frac{\partial \Psi(\vec r,t)}{\partial t}  = \Big\{  -\sum^{M_T}_{a=1} \frac{\hbar^2}{2m^*}\nabla^2_a  +  U(\vec r,t) \Big\} \Psi(\vec r,t),
\end{equation}
where for simplicity we have considered a solid-state system where the lattice-electron interaction is approximately included into the electron effective mass, $m^*$. The term $U(\vec r,t)$ is the potential energy that, here, defines the Coulomb interaction among electrons.
From Eq. (\ref{Scho}) it can be demonstrated that the probability distribution, $|\Psi(\vec r,t)|^2$, obeys  the following continuity equation:
\begin{equation}
\frac{\partial |\Psi(\vec r,t)|^2}{\partial t}  +  \sum^{M_T}_{a=1}  \nabla_a \vec j_a(\vec r,t) = 0,
\label{Continuity}
\end{equation}
where $\vec j_a(\vec r,t)$ is the $a-$th component of the usual probability current density \cite{o.cohen1978book}.
From Eq. (\ref{Continuity}), the vector field defined as:
\begin{equation} \label{Velocity}
\vec v_a(\vec r,t)  = \frac {{\vec j_a(\vec r,t)} } {{|\Psi(\vec r,t)|^2}},
\end{equation}
can be interpreted as a \emph{velocity field} for the particle $a$. This velocity can be used to define trajectories in the configuration space:
\begin{equation}
\vec r_a^\alpha(t) = \vec r_a^\alpha(t_o) + \int\limits_{t_o}^{t} \vec v_a(\vec r^\alpha(t'), t') dt',
\label{Newtonlike}%
\end{equation}
with $\vec r^\alpha(t) = \{ \vec r_1^\alpha(t),...,\vec r_{M_T}^\alpha(t) \}$. The superindex $\alpha$ takes into account the uncertainity associated with the initial quantum state $\Psi(\vec r,t_o)$, and it is defined through the so-called \emph{Quantum equilibrium condition} \cite{o.durr2004equilibrium,o.oriols2011book} at the initial time $t_o$:
\begin{equation} \label{Equil_hypo}
|\Psi(\vec r,t_o)|^2 = \mathop {\lim }\limits_{M_\alpha \to \infty } \frac{1}{M_\alpha}\sum\limits_{\alpha = 1}^{M_\alpha} \prod_{a=1}^{M_T} \delta(\vec r_a - \vec r_a^{\alpha}(t_o)).
\end{equation}
It can be easily demonstrated that the evolution of the above infinite set of quantum trajectories, $\alpha=1,2,...,M_\alpha \rightarrow \infty$, reproduces the probability distribution, $|\Psi(\vec r,t)|^2$, at any time.
Equations (\ref{Scho}), (\ref{Velocity}), (\ref{Newtonlike}) and (\ref{Equil_hypo}) constitute a basic set of equations describing Bohmian mechanics \cite{o.bell2004book,o.durr2012book,o.durr2009book,o.oriols2011book}. As will be discussed in \sref{Bohm2}, in Bohmian mechanics, any observable, for example the current $I$, is a function of the particle trajectories $\vec r^\alpha(t)$ in \eref{Newtonlike}, i.e. $I(\vec r^\alpha(t))$. It is in this regard that Bohmian mechanics is a trajectory-based formulation of quantum mechanics.

\subsection{Bohmian explanation of the measurement process}
\label{Bohm2}

The Bohmian explanation of the quantum measurement is, perhaps, the most attractive (and also ignored) feature of the Bohmian explanation of the quantum nature \cite{o.bell1990pw,o.bell2004book,o.durr2012book,o.oriols2011book}. Although the Bohmian and the orthodox explanations of a measurement produce the same probabilistic predictions, the mathematical implementation of the equations of motion in each case is quite different.

\begin{figure}
\includegraphics[width=0.57\columnwidth]{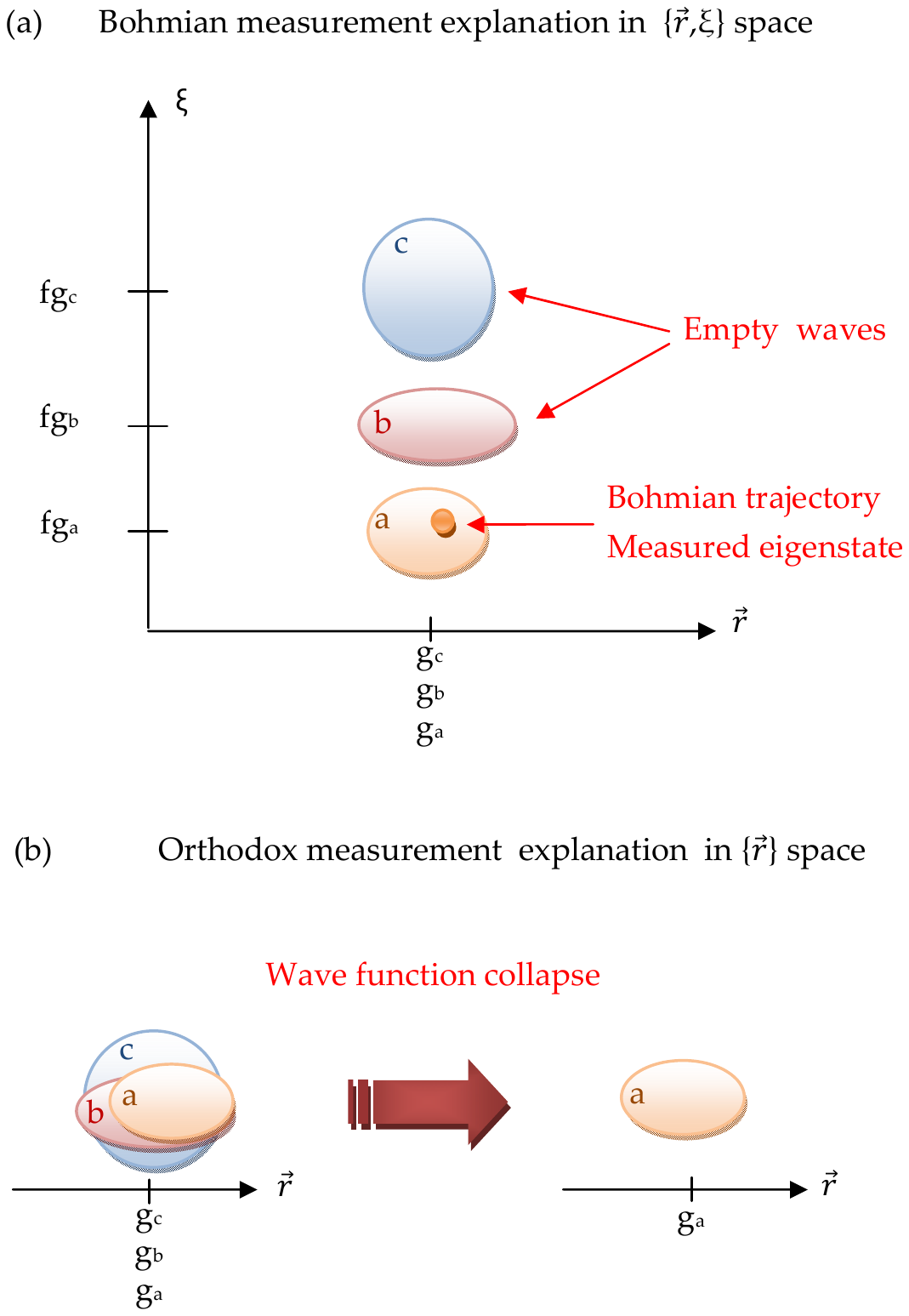}
\centering
\caption{(Color online) (a) Bohmian measurement in the $\{\vec r,\xi\}$ configuration space: from the nonoverlapping many-particle (system + apparatus) wave function, only the $g_a$
part of the wave function where the Bohmian trajectory is present is needed to compute the evolution of the Bohmian system. (b) Orthodox
measurement in the $\{\vec r\}$ space: the (system) wave function collapses into the $g_a$ part when the measurement takes place.}  
\label{o.figure2}
\end{figure}

In the standard interpretation of quantum theory, the projective measurement process is defined in a particular quantum region, the system. See \fref{o.figure2}(b). The state of the quantum system in this particular region is determined through the wave function $\psi_S(\vec r,t)$. The process of measuring a particular magnitude is mathematically defined through an operator, for example $\hat{G}$, acting on the wave function. The possible outcomes of the measurement process correspond to one of the possible eigenvalues $g$ of this operator that satisfy the equation $\hat{G}\psi_g(\vec r) = g\psi_g(\vec r)$, with $\psi_g(\vec r)$ being an eigenvector of this operator. The set $\psi_g(\vec r)$ forms an orthonormal basis of the Hilbert space of the quantum system so that the wave function at the initial time $t$ can be written as:
\begin{equation}
\label{om.bmeasure1}
\psi_S(\vec r,t) = \sum_{g} c_g(t)\psi_g(\vec r),
\end{equation}
with $c_g(t)$ a complex value with the only restriction that $\sum_{g} |c_g(t)|^2 = 1$, which ensures that $\psi_S(\vec r,t)$ is well normalized. When measuring the eigenvalue $g_a$ the total wave function, $\psi_S(\vec r,t)$ collapses into $\psi_{g_a}(\vec r)$. Then, the probability of getting the value $g_a$ in the measuring apparatus is just $P_{g_a} = |c_{g_a}(t)|^2$. In order to avoid unnecessary complications we have assumed that this basis has no degeneracy.

In order to mathematically define the measurement process in the Bohmian formalism, apart from the degrees of freedom $\vec r$ of the system, the degrees of freedom of the positions of the pointer $\vec \xi$ belonging to the measuring apparatus are required \cite{o.Holand1993,o.bohm1993book,o.bell2004book,o.durr2012book,o.oriols2011book}. 
Thus, we define a total wave function $\Phi(\vec r,\vec \xi,t)$ in a larger configuration space that
includes the system plus the measuring apparatus, $\{\vec r,\vec \xi\}$. According to the Bohmian postulates, we select a particular trajectory $\{\vec r^\alpha(t),\vec \xi^\alpha(t)\}$ of this larger configuration space. Then, in order to say that a measuring apparatus is able to correctly determine the eigenvalues $g$, there are some necessary conditions that the entire system has to satisfy.

First, the pointer positions $\vec \xi^\alpha(t)$ of such an apparatus have to be restricted to a particular region, $\vec \xi^\alpha(t)\in S_{g_1}$, every
time that the quantum system is in the eigenstate $\psi_{g_1}(\vec r)$. We define $S_{g}$ as a restricted set of positions in the space of the ammeter position $\xi$. Let us define $\Phi_{g_1}(\vec r,\vec \xi,t)$ as the total wave function that fits with the property that any experiment whose
quantum system is described by $\psi_{g_1}(\vec r)$ implies that the pointer points in the particular region, $\vec \xi^\alpha(t)\in S_{g_1}$.

Second, the subspaces $S_{g1}$ and $S_{g2}$ of the whole configuration space must be non-overlaping during the measurement, i.e. $S_{g1} \cap S_{g2} = 0$. We have defined the restricted region allowed by the pointer positions associated with a second eigenstate, $\psi_{g_2}(\vec r)$, as $\vec \xi^\alpha(t)\in S_{g_2}$. This implies that the states $\Phi_{g_1}(\vec r,\vec \xi,t)$ and $\Phi_{g_2}(\vec r,\vec \xi,t)$ do not overlap in the configuration space during the measurement.

Thus, given a ``good'' (projective) measurement apparatus, and given that the eigenstates $\psi_g(\vec r)$ form a complete basis, during the measurement, the
only good decomposition for $\Phi_g(\vec r,\vec \xi,t)$ is:
\begin{equation}
\Phi_g(\vec r,\vec \xi,t) = f_g(\vec \xi,t) \; \psi_g(\vec r).
\label{om.bmeasure2}
\end{equation}
We emphasize that $ f_g(\vec \xi,t)$ is a normalized function because $\Phi_g(\vec r,\vec \xi,t)$ and $\psi_g(\vec r)$ are also normalized functions in their respective configuration spaces. By construction, $f_{g_1}(\vec \xi,t) \cap f_{g_2}(\vec \xi,t) = 0$ during the measuring time. Thus, even if $\psi_{g_1}(\vec r)$ and $\psi_{g_2}(\vec r)$ overlap, the states $\Phi_{g_1}(\vec r,\vec \xi,t)$ and $\Phi_{g_2}(\vec r,\vec \xi,t)$ do not overlap in the larger configuration space. See \fref{o.figure2}(a). We can then ensure that an arbitrary wave function of the quantum system, \eref{om.bmeasure1}, can be rewritten in the whole configuration space associated to a good measuring apparatus as:
\begin{equation}
\Phi(\vec r,\vec \xi,t) = \sum_{g} c_g(t) \;\; f_g(\vec \xi,t) \; \psi_g(\vec r).
\label{om.bmeasure3}
\end{equation}
In summary, during the measurement, the only total wave functions that can \textit{live} in the entire quantum system that includes a good measuring apparatus of the eigenvalues $g$ are the ones written in \eref{om.bmeasure3}. An example of such wave functions is depicted in \fref{o.figure2}(a). It is important to notice that \eref{om.bmeasure3} implies no restriction on the wave
function $\psi_S(\vec r,t)$ but only on the total wave function $\Phi(\vec r,\vec \xi,t)$. If these restrictions are not respected, we can find other types of total wave functions in the configuration space $\{\vec r,\vec \xi\}$, but they would be incompatible with stating that we have an apparatus that is able to measure the eigenavalue $g$ with certainty at time $t$.

We can now show quite trivially how the projective measurement is exactly reproduced within Bohmian mechanics. As we have mentioned, apart from the wave function \eref{om.bmeasure3}, we have to select an initial trajectory $\{\vec r^\alpha(0),\vec \xi^\alpha(0)\}$. Such a trajectory will evolve driven by the total wave function, and during the measurement, the particle trajectory $\{\vec r^\alpha(t),\vec \xi^\alpha(t)\}$ will be situated in only one of the nonoverlapping wave packets of \eref{om.bmeasure3}, for example $f_{g_a}(\vec \xi,t) \psi_{g_a}(\vec r)$ as depicted in \fref{o.figure2}a. Thus, the pointer positions will be situated in $\vec \xi(t)\in S_{g_a}$ and we will conclude with certainty that the eigenvalue of the quantum system is $g_a$. In addition, the subsequent evolution of this trajectory can be computed from $f_{g_a}(\vec \xi,t) \psi_{g_a}(\vec r)$ alone. In other words, we do not need the entire wave function \eref{om.bmeasure3} because the particle velocity can be computed from $f_{g_a}(\vec \xi,t) \psi_{g_a}(\vec r)$. The rest of circles of \fref{o.figure2}(a) are empty waves that do not overlap with $f_{g_a}(\vec \xi,t) \psi_{g_a}(\vec r)$ so that they have no effect on the velocity of the Bohmian particle. This is how the orthodox collapse is interpreted within Bohmian mechanics. Let us mention that we have only considered projective measurements. Other types of measurements are also possible, which do not collapse the wave function into an eigenstate. Such measurements can also be explained within Bohmian mechanics with an extension of the ideas discussed here \cite{o.bohm1993book,o.durr2004equilibrium,o.durr2012book}. A simple numerical example about the quantum measurement of the total (displacement plus particle) current using Bohmian mechanics will be discussed in \sref{Bitlles1}.

A final remark about the quantum measurement is necessary. The Bohmian measurement process explained above implies increasing the number of degrees of freedom that one has to simulate from $\{\vec r\}$ to $\{\vec r,\vec \xi\}$. Sometimes, then, the use of Hermitian operators acting only on the wave function of the quantum system with the ability of providing the outcomes of the measurement process without the explicit simulation of the measuring apparatus is very welcomed. The reader is referred to Appendix \ref{appendixC} for a detailed discussion on how to include operators in Bohmian mechanics. Let us emphasize, however, that we are talking only at the computational level. In simple words, operators are not needed in Bohmian mechanics, but they are a very helpful mathematical tool in practical computations. These ideas are emphasized by Zangh\`i, Goldstein, D\"{u}rr and coworkers when they talk about the ``naive realism about operators'' \cite{o.durr1997naive,o.durr2004equilibrium,o.durr2012book}.

\subsection{Bohmian mechanics for many-particle systems}
\label{Bohm3}

The active region of the electron device shown in \fref{o.figure4} can contain hundreds of electrons. However, as we mentioned, the many-particle Schr\"{o}dinger equation in \eref{Scho} can be solved only for very few degrees of freedom. A standard way to proceed consists then on reducing the complexity of the problem by \emph{tracing out} certain degrees of freedom. This process ends up with what is called the \emph{reduced density matrix}. When the reduced density matrix is used, its equation of motion is no longer described by the Schr\"odinger equation but in general by a non-unitary operator. The reduced density matrix is no longer a pure state, but a mixture of states and its evolution is in general irreversible \cite{o.DiVentra2008book}. In this section we discuss how Bohmian mechanics allows us to reduce the complexity in a very different way. As it will be shown below, the concept of \emph{conditional wave function} \cite{o.durr2004equilibrium} provides an original tool to deal with many-body open quantum systems \cite{o.oriols2007prl,o.durr2005fp,PRL_Alb}.

\subsubsection{The Conditional Wave Function}
\label{CWF}

Consider a bipartite quantum system $A+B$ whose spatial coordinates can be split as $\vec r = \{\vec r_a, \vec r_b\}$. We define $\vec r_a$ as the position of the $a-$electron in $\mathbf{R}^3$, while $\vec r_b=\{\vec r_1, ....,\vec r_{a-1},\vec r_{a+1},....,\vec r_{M_T}\}$ are the positions of the rest of electrons in a $\mathbf{R}^{3 (M_T-1)}$ space. The actual particle trajectories are accordingly denoted by $\vec r(t) = \{\vec r^\alpha_a(t), \vec r^\alpha_b(t)\}$. \emph{How can one assign a wave function to the system $A$?} In general this is not possible if the two subsystems are entangled, i.e. the total wave function cannot be written as a product $\Psi(\vec r) = \psi_a(\vec r_a)\psi_b(\vec r_b)$. However, we can modify our question and ask what is the wave function of the subsystem $A$ that provides the exact velocity $\vec v_a$ given a particular position $\vec r^\alpha_b(t)$ in $B$. The answer given by Bohmian mechanics is the so called \emph{conditional wave function} \cite{o.durr2004equilibrium,o.durr2005fp}:
\begin{equation}
\phi_{a}(\vec r_a,t) = \Psi(\vec r_a,\vec r^\alpha_b(t),t),
\label{Conditional}
\end{equation}
which constitutes a multi-dimensional slice of the whole wave function. In \eref{Conditional} we omit (for simplicity) the dependence of each conditional wave function $\phi_{a}(\vec r_a,t)$ on $\alpha$.

In order to use the conditional wave function to reduce the degrees of freedom of a system, we must know how it evolves in time. It can be demonstrated \cite{o.oriols2007prl} that $\phi_{a}(\vec{r}_a, t)$ obeys the following wave equation:
\begin{eqnarray}\label{Conditional_eq}
i\hbar\frac{\partial \phi_a(\vec r_a,t)}{\partial t}=\Big\{-\frac{\hbar^2}{2m} \nabla^2_a+U_{a}(\vec r_a,\vec r^\alpha_b(t),t) \nonumber \\ +G_{a}(\vec r_a,\vec r^\alpha_b(t),t)+ i J_{a}(\vec r_a,\vec r^\alpha_b(t),t) \Big\} \phi_{a}(\vec r_a,t).
\end{eqnarray}
The explicit expressions of the potentials $G_{a}(\vec r_a,\vec r^\alpha_b(t),t)$ and $J_{a}(\vec r_a,\vec r^\alpha_b(t),t)$ that appear in (\ref{Conditional_eq}) can be found in Ref. \cite{o.oriols2007prl}, however, their numerical values are in principle unknown and need some educated conjectures. On the other hand, the term $U_{a}(\vec r_a,\vec r^\alpha_b(t),t)$ can be any type of many-particle potential defined in the position-representation, in particular, it can include short-range and long-range Coulomb interactions. For simplicity, we have divided  the total electrostatic potential energy among the $M_T$ electrons that appears in (\ref{Scho}), into two parts $U(\vec r_a,\vec r^\alpha_b(t),t) = U_{a}(\vec r_{a},\vec r^\alpha_{b}(t),t) + U_{b}(\vec r^\alpha_{b}(t),t)$. The remaining term $U_{b}(\vec r^\alpha_{b}(t),t)$ that do not involve the variable $\vec r_a$ is contained in the coupling potential $G_a$ in (\ref{Conditional_eq}).

From a practical point of view, all quantum trajectories $\vec r^\alpha(t)=\{\vec r_1^\alpha(t), \vec r_2^\alpha(t),..,\vec r_{M_T}^\alpha(t)\}$ have to be computed simultaneously. In order to gather all the above concepts, let us discuss an hypothetical computation with conditional wave functions by detailing a sequential procedure:
\begin{enumerate}
\item At the initial time $t_o$, we fix the initial position of all $a=1,...,M_T$ particles, $\vec r^\alpha_a(t_o)$, according to (\ref{Equil_hypo}), and their associated single-particle wave function $\phi_{a}(\vec{r}_a, t_o)$. We define this set of positions with the superindex $\alpha=1$.
\item From all particle positions, we compute the exact value of the potential $U_{a}(\vec{r}_{a},\vec{r}^\alpha_{b}(t_o),t)$ for each particle. An approximation for the terms $G_{a}$ and $J_{a}$ is required at this point.
\item We then solve each single-particle Schr\"{o}dinger equation, (\ref{Conditional_eq}), from $t_o$ till $t_o+dt$.
\item From the knowledge of the single-particle wave function $\phi_{a}^{\alpha}(\vec{r}_a, t_o+dt)$, we can compute the velocities $\vec{v}_{a}^\alpha(t_o+dt)$ for each $a$-particle.
\item With the previous velocity, we compute the new position of each $a$-particle as $\vec{r}^\alpha_{a}(t_o+dt)=\vec{r}^\alpha_{a}(t_o)+\vec{v}_{a}^\alpha(t_o+dt)dt$.
\item Finally, with the set of new positions and wave functions, we repeat the whole procedure (steps 2 till 5) for another infinitesimal time $dt$ till the total simulation time is finished.
\end{enumerate}
Another experiment (or the same experiment at another time) will require selecting different initial positions in step 1 and repeating the whole loop. The advantage of the above algorithm using (\ref{Conditional_eq}) instead of (\ref{Scho}) is that, in order to find approximate trajectories, $\vec r^\alpha_a(t)$, we do not need to evaluate the wave function and potential energies in the whole configuration space in \eref{Scho}, but only over a smaller number of configuration points, $\{\vec r_a, \vec r^\alpha_b(t)\}$, associated with those trajectories defining the highest probabilities in (\ref{Equil_hypo}).

For spinless electrons, the exchange interaction is naturally included in (\ref{Conditional_eq}) through the terms $G_{a}$ and $J_a$. Due to the Pauli exclusion principle, the modulus of the wave function tends to zero, $R(\vec r_a,\vec r^\alpha_b(t),t)\to 0$, in any neighborhood of $\vec r_{a_j}$ such that $|\vec r_{a_j}-\vec r^\alpha_{b_k}(t)|\to 0$ with $j$ and $k$ referring to the individual particles of systems $A$ and $B$ respectively. Thus, both terms, $G_{a}(\vec r_a,\vec r^\alpha_b(t),t)$ and $J_{a}(\vec{r}_{a},\vec r^\alpha_b(t),t)$, have asymptotes at ${\vec{r}_{a_j}}\to {\vec {r}^\alpha_{b_k}}(t)$ that \textit{repel} the $a-$ particle from other electrons. However, in order to exactly compute the terms $G_a$ and $J_a$ we must know the total wave function, which is in principle unknown. There are however a few ways to introduce the symmetry of the wave function without dealing directly with these two coupling terms \cite{o.oriols2007prl,o.alarcon2013pcm}.

\subsubsection{An example: numerical results for a non-separable potential}
\label{example}
In order to numerically show the ability of the conditional wave functions discussed above to treat many-particle systems, we apply the above algorithm to a simple two-electrons system under a non-separable harmonic Hamiltonian. We consider two 1D particles so that the configuration space is $\mathbf{R}^2$. The object described by Eq. (\ref{Conditional}) can be easily understood in this simple case. Here the conditional wave function $\phi_1( x_1,t)$ would represent a 1D slice of the whole 2D wave function centered on a particular configuration point of $x_2$, i.e. $\phi_{1}(x_1,t) = \Psi(x_1,x_2^\alpha(t),t)$. We use a non-separable potential energy:
\begin{equation}
U(x_1,x_2) = F \cdot (x_1 - x_2)^2,
\end{equation}
with $F = 10^{12} eV/m^2$ quantifying the strength of the many-body interaction. The many-body wave function $\Psi(x_1, x_2, t)$ can be solved exactly from Eq. \eref{Scho} with $M_T = 2$. Once the exact 2D wave function $\Psi(x_1, x_2, t)$ is known, we can compute the exact 2D Bohmian trajectories straightforwardly from \eref{Velocity}, \eref{Newtonlike} and \eref{Equil_hypo}. The initial wave function is a direct product, ${\psi}_{1}(x_1,0) \cdot {\psi}_{2}(x_2,0)$ of two Gaussian wave packets as the one defined in \eref{om.finite-difference_innitial} of Appendix \ref{appendixA}. In particular, we consider $E_{o1}=0.06$ eV, $x_{c1}=-50$ nm and $\sigma_{x1}=25$ nm for the first wave packet, and $E_{o2}=0.04$ eV, $x_{c2}=50$ nm and $\sigma_{x2}=25$ nm for the second. In this particular example, the exchange interaction among electrons has been disregarded. For similar example with exchange interaction see \cite{o.alarcon2013pcm}.

\begin{figure}[ht]%
\centering
\includegraphics[width=0.57\columnwidth]{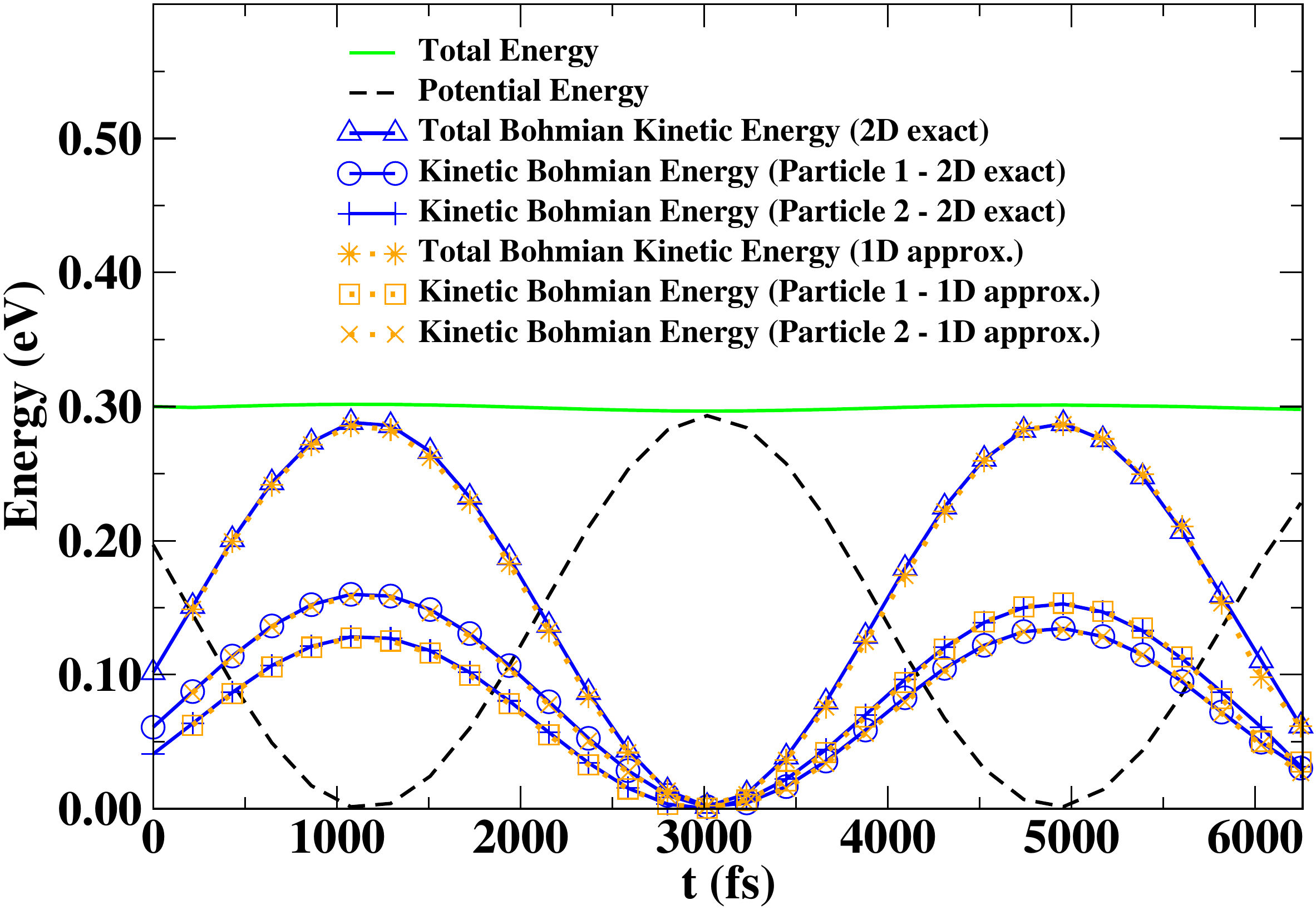}
\caption
{%
\footnotesize{(Color online) Time evolution of individual (ensemble averaged) Bohmian kinetic energies of identical two-electron system without exchange interaction under a non-separable potential computed from $2D$ exact and $1D$ approximate solutions. }}
\label{o.figure3}
\end{figure}

In \fref{o.figure3}, we have plotted the ensemble (Bohmian) kinetic energy for the two electrons $a=1$ and $a=2$ using a set of $M_\alpha=160000$ Bohmian trajectories. 
The exact expression of the Bohmian kinetic energy
is discussed in Appendix \ref{appendixC} and defined in \eref{kineticbohm}. We first compute the results directly from the 2D exact wave function solution of \eref{Scho}. We emphasize that there is an interchange of kinetic energies between the first and second particle (see their kinetic energy in the first and second oscillation) indicating the many-particle nature of the system. This effect clearly manifests that the Hamiltonian of that quantum system is non-separable. Alternatively, we can compute the Bohmian trajectories without knowing the many-particle wave function, i.e. using the conditional wave function $\phi_a(x_a,t)$ solution of \eref{Conditional_eq} with a proper approximation of $G_{a}$ and $J_{a}$. Here, we consider a zero order Taylor expansion around $x_a^\alpha(t)$ for the unknown potentials $G_{a}$ and $J_{a}$.  In other words, we consider them as purely time-dependent potentials, $G_{a}(x_{a},x^\alpha_{b}(t),t) \approx G_{a}^{''}(x^\alpha_a(t),x^\alpha_{b}(t),t)$ and identically $J_{a}(x_{a},x^\alpha_{b}(t),t) \approx J_{a}^{''}(x^\alpha_a(t),x^\alpha_{b}(t),t)$. This constitutes the simplest approximation. Then, we know that these purely time-dependent terms only introduce a (complex) purely time-dependent phase in the solution of \eref{Conditional_eq}, so we can write $\phi_a(x_a,t)$ as:
\begin{equation}
\phi_a(x_a,t)= \tilde{\psi}_{a}(x_a,t) \exp (z^\alpha_{a}(t)),
\label{mpnocoulomb}
\end{equation}
where the term $z^\alpha_{a}(t)$ is a (complex) purely time-dependent phase that has no effect on the trajectory $x^\alpha_a(t)$. Under the previous approximation, \eref{Conditional_eq} can be simplified into the following equation for the computation of $\tilde{\psi}_{a}(x_a,t)$:
\begin{equation}
i\hbar\frac{\partial \tilde{\psi}_{a}(x_a,t)}{\partial t} = \Big( -\frac{\hbar^2}{2m}\frac{\partial^2}{\partial {x^2_a}} + U_a (x_a, x^\alpha_b(t)) \Big) \tilde{\psi}_{a}(x_a,t),
\label{setpseudo}
\end{equation}
where the potential energies are $U_1 (x_1, x^\alpha_2(t))=F(x_1-x^\alpha_2(t))$ for $a=1$ and $U_2 (x_2, x^\alpha_1(t))=F(x^\alpha_1(t)-x_2)$ for $a=2$. For this particular scenario, our simplest approximation for the unknown terms $G_{a}$ and $J_{a}$ works perfectly and the agreement between 2D exact results and our 1D approximation is excellent (see \fref{o.figure3}). We have computed the ensemble energies in order to justify that the algorithm is accurate not only for an arbitrarily selected set of Bohmian trajectories, but for an ensemble of them. We use the steps 1 till 6 explained above for each trajectory $\{x_1^\alpha(t),x_2^\alpha(t)\}$. Each trajectory is also computed from a 2D version of the algorithm explained in Appendix \ref{appendixA}.

An improvement over the simple approximation used here for  $G_{a}$ and $J_{a}$, when constructing the conditional (Bohmian) wave functions, is necessary in other types of interacting potentials to get the same degree of accuracy as shown in Fig. \ref{o.figure3}. A possibility that will be explored in future works is following the ideas presented in \cite{o.travis2010fp,Synthese} where a full (infinite) set of equations for an exact description of the conditional wave functions is presented.

\section{The BITLLES Simulator: Time-resolved Electron Transport}
\label{Bitlles}
The preceding section was devoted to discuss the ability of Bohmian mechanics to provide, in one hand, a simple explanation of the measurement process, and on the other hand, an algorithm to approximate the many-body problem. Now, we will focus on the application of this Bohmian machinery to build up a quantum electron devices simulator called BITLLES \footnote{BITLLES is the acronym of Bohmian Interacting Transport for non-equiLibrium eLEctronic Structures. See the website {http://europe.uab.es/bitlles.}}. We have divided this section into three parts.

In \sref{Bitlles1} we argue in detail how the effects of the measuring apparatus can be taken into account in the evolution of the active region of electronic devices. As a particular case of the measurements that can be made over an electron device, we focus here on the electrical current. By means of a simple model, we find a relationship between the current measured on the ammeter and the Bohmian trajectories of the system. By solving the system-apparatus Schr\"odinger equation for this simplified scenario, we argue that whenever the back action of the apparatus on the system trajectories is not much relevant, we can avoid the explicit computation of the pointer degrees of freedom and focus only on the system to compute the electrical current (without invoking any non-unitary evolution). \sref{Bitlles2} is devoted to explain how to go beyond mean-field approaches and include full Coulomb correlations. We demonstrate that Coulomb interactions can be effectively included by defining a Poisson equation for each carrier and a complet set of time-dependent boundary conditions. We argue on the crucial role played by the leads in assuring overall-charge neutrality and current conservation. Finally, in \sref{Bitlles3} we summarize the main pieces that define an electron injection model valid for systems with and without electron confinement. From a practical point of view, this model introduces, apart from the uncertainty in the initial position of the quantum trajectories, an additional randomness on the properties of the injected electrons related to their energies, velocities, etc. It is in this regard that the BITLLES simulator can be somehow understood as a quantum Monte Carlo algorithm. All the results presented in the following sections have been computed using BITLLES. 
It is worth noticing here that, because of the many analogies between the semi-classical and the Bohmian descriptions (both in terms of trajectories) of electron transport, 
the BITLLES simulator also includes a semiclassical limit which corresponds to a many-particle version of the well known Monte Carlo solution of the Boltzmann equation \cite{JAP,intech,JSTAT,APL,IJNM,Albareda2008}.

\subsection{On the role of the measuring apparatus}
\label{Bitlles1}

\begin{figure}
\centering
\includegraphics*[width=0.57\columnwidth]{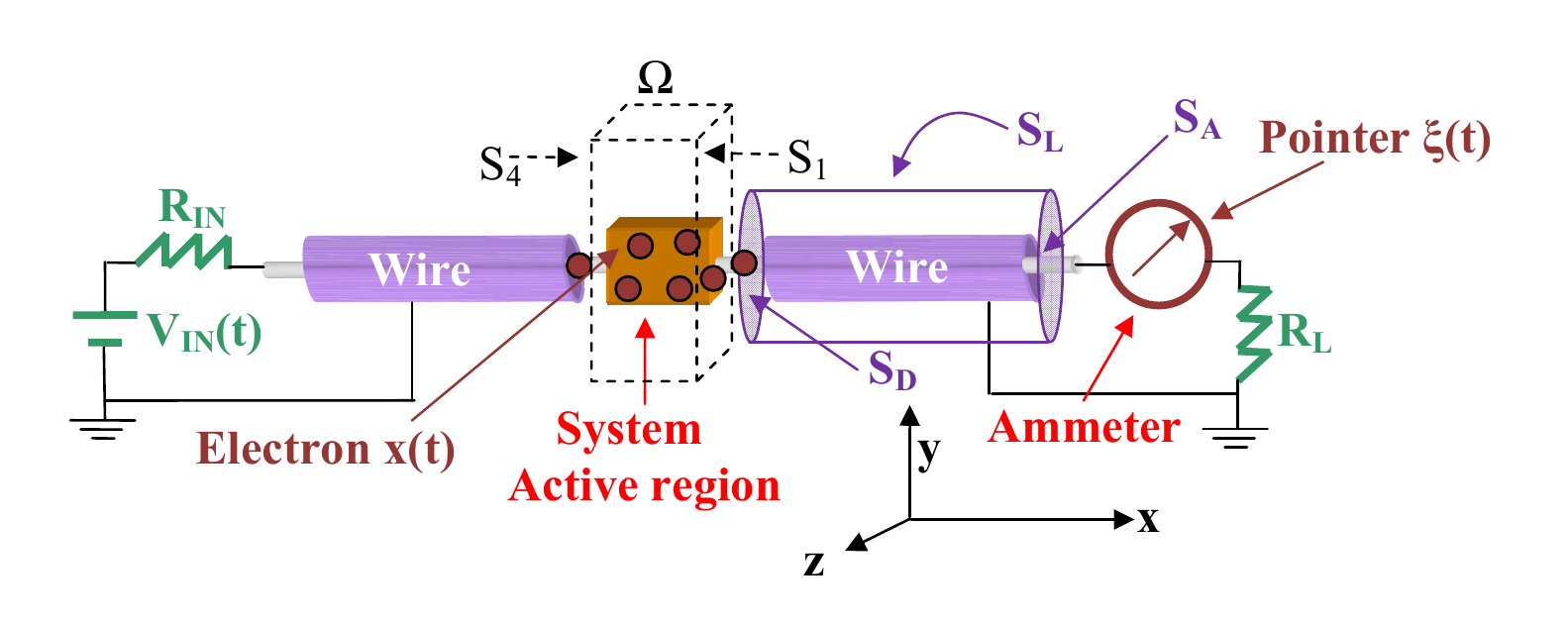}
\caption{ (Color online) Schematic representation of a typical electrical circuit used in this chapter for studying the current measurement in electrical device. Device simulators compute the current on the surface, $S_{D}$, of the active region, while the ammeter measures it on the surface,  $S_{A}$. }
\label{o.figure4}
\end{figure}

The functionality of any electronic device is determined by the relationship between the current measured by an ammeter and the voltage imposed at the external battery (see \fref{o.figure4} for a description of a typical electronic circuit). Any measurement of a classical or quantum device implies an interaction between the apparatus and the measured system.

As discussed in \sref{Intro1}, the \textit{orthodox} time-evolution of the wave function $\Psi(\vec r,t)$ is governed by two different dynamical laws. First, there is a dynamical (deterministic) evolution according to \eref{Scho} when the system is not measured. Second, there is an (stochastic) evolution known as \textit{collapse} of the system wave function when it interacts with a measurement apparatus. On the contrary, Bohmian mechanics does not differentiate between measuring and non-measuring evolutions \cite{o.durr2012book,o.bell2004book,o.oriols2011book}. In particular, we define $\vec \xi=\{\vec \xi_1,...,\vec \xi_K\}$ as the $K$ degrees of freedom that conform the pointer of a measuring apparatus (for example, the ammeter). Because of the pointer, we have to deal with a wave function $\Phi(\vec r,\vec \xi,t)$ whose equation is equivalent to (\ref{Scho}) but in the extended configuration space. The pointer position $\vec \xi^\alpha(t)$ and system position $\vec r^\alpha(t)$ move according to their equations of motion equivalent to \eref{Newtonlike}. The (Bohmian) position of a \emph{good} pointer $\vec \xi^\alpha(t)$ are supposed to be correlated with the system position $\vec r^\alpha(t)$. The stochastic nature of the quantum measurement is recovered here because in the experimental setup we do not know the initial Bohmian positions $\vec \xi^\alpha(t_o)$ and $\vec r^\alpha(t_o)$, which have to be selected in the simulation according to the quantum equilibrium condition \eref{Equil_hypo}.
The Bohmian explanation of the measurement process has, however, two technical difficulties:

Firstly, we have to specify which
Hamiltonian determines the evolution of the system plus apparatus. This difficulty is similar to
specifying which operator provides good information about the measuring process in the \emph{orthodox}
quantum mechanics \cite{o.durr2004equilibrium}.

The second difficulty is related to the computational limitations while solving the many-particle
Schrödinger equation. The Schrödinger equation with the addition of the pointer is most of the times unsolvable. This technical difficulty is non existent when using operators because they act only on the system's wave function.

\subsubsection{The system plus apparatus Schr\"{o}dinger equation}
\label{theory_measure}

The idea of including the pointer as an additional degree of freedom in the Schr\"{o}dinger equation was already proposed by von Neumann in 1932 within  \textit{orthodox} quantum mechanics, when trying to provide a macroscopic explanation of the \textit{collapse} of the wave function \cite{o.vonNeumann}. Let us consider, for the moment, only two degrees of freedom. The variable $\vec r_1$ for the system and $\xi$ for the center of mass of the apparatus (in a 1D system). The interaction among them is determined by the following Hamiltonian:
\begin{eqnarray}
H_{int}=-i \hbar\lambda A(\vec r_1) \frac {\partial } {\partial \xi},
\label{o.meq1}
\end{eqnarray}
where $A(\vec r_1)$ is the magnitude of the system that we want to measure and $\lambda$ is a coupling constant. From Eq. (\ref{o.meq1}), the $\xi-$component of the \textit{local} momentum (velocity) of the wave function $\Phi(\vec r_1,\xi,t)$ will depend on the magnitude of $A(\vec r_1)$. Therefore, an initially localized wave packet $\Phi(\vec r_1,\xi,0)$ will spread in the $\xi-$direction because of the different $\xi-$velocities. Then, the only new ingredient that we have to include when dealing with Bohmian mechanics is the presence of the pointer and system (Bohmian) trajectories, $\{\vec r_1^\alpha(t),\xi^\alpha(t)\}$.  Next, we specify what expression we have to adopt for $A(\vec r_1)$ in order to effectively compute the electrical current \cite{o.albareda2013jcel}.

It is common to compute the electrical current on the (simulated) surface $S_D$ of \fref{o.figure4}, while a real measurement is performed on the (non-simulated) surface $S_A$. It is then crucial to understand in which extension is the current on $S_A$ equal to that on $S_D$. In fact, these currents will be only equal if we consider the \emph{total current} $I(t)=I_p(t)+I_d(t)$, where $I_p(t)$ and $I_d(t)$ are respectively the particle and displacement components discussed in \sref{Intro2}. Since the Maxwell equations ensure that the total current density  $\vec J(\vec r_1,t)$ is a vector with a null divergence, then we can write $\int_{S} \vec J(\vec r_1,t) d\vec s=0$ for a closed surface $S=\{S_D,S_A,S_L\}$ where $S_L$ is a the surface parallel to the transport direction in the cable as drawn in \fref{o.figure4}. In particular, for a cable we can assume $\int_{S_L} \vec J(\vec r_1,t) d\vec s=0$, so we finally get  $\int_{S_D} \vec J(\vec r_1,t) d\vec s=-\int_{S_A} \vec J(\vec r_1,t) d\vec s$.

The function $A(\vec r_1)$ in Eq. (\ref{o.meq1}) has then to be related to the total current $I(t)$ of the system, meaning that $I(t)$ has to be somehow linked to the positions $\vec r_1^\alpha(t)$ of the particles of the system. The total current measured by an ammeter, $I^\alpha(t)$, for a particular trajectory $\vec r_1^\alpha(t)=\{x^\alpha(t),y^\alpha(t),z^\alpha(t)\}$, can be defined as the time-derivative of the following \textit{particle} plus \textit{displacement} charges:
\begin{equation}
I^\alpha(t)=\frac {d \left( Q_{p}(\vec r_1^\alpha(t))+Q_{d}(\vec r_1^\alpha(t)) \right) } {dt} \equiv \frac {d Q \left(\vec r_1^\alpha(t) \right) } {dt},
\label{o.meq2}
\end{equation}
where we define the \textit{conduction} charge as:
\begin{equation}
Q_{p}(\vec r_1)=-q \int_{S_D} dy\rq{} \; dz\rq{} \int_{x\rq{}=x_{D}}^{\infty}dx \rq{} \delta (\vec r_1\rq{}-\vec r_1),
\label{o.meq3}
\end{equation}
$q$ being the (unsigned) electron charge and $x_{D}$ the x-position of the lateral surface $S_{D}$. The particle charge is $Q_{p}(\vec r_1)=-q$ only if the electron is located at the right of $x_D$ and inside the lateral surface $S_D$. Identically, we can interpret the \textit{displacement} charge in \eref{o.meq2}, as:
\begin{equation}
Q_{d}(\vec r_1)=\int_{S_{D}} \varepsilon(\vec r_1\rq{}) \vec E(x,y,z,x_{D},y\rq{},z\rq{})d\vec s,
\label{o.meq4}
\end{equation}
being $\vec E(x,y,z,x_{D},y\rq{},z\rq{})$ the electric field generated at $\vec r_1\rq{}=\{x_{D},y\rq{},z\rq{}\}$ of the surface $S_D$ by one electron at $\vec r_1=\{x,y,z\}$. It is important to emphasize that $Q_{d}(\vec r_1)$ is different from zero independently of the distance between the electron and the surface. The generalization of $Q_{p}(\vec r_1)$ and $Q_{d}(\vec r_1)$ to an arbitrary number of electrons $Q_{p}(\vec r_1,..,\vec r_{M_T})$ and $Q_{d}(\vec r_1,..,\vec r_{M_T})$ is quite simple.  Finally, in our ammeter model, the von Neumann term $A$ in Eq. (\ref{o.meq1}) is the \textit{conduction} plus \textit{displacement} charges defined in Eqs. (\ref{o.meq3}) and (\ref{o.meq4}) divided by the elementary charge $q$, i.e.:
\begin{equation}
A(\vec r_1,..,\vec r_{M_T})=-\frac{Q(\vec r_1,..,\vec r_{M_T}) }{q},
\label{o.meq5}
\end{equation}
where the (irrelevant) minus sign appears just to provide a positive pointer movement when an electron moves from left to right in \fref{o.figure4}, which corresponds to a negative net current. Certainly, other modeling of the ammeter are possible, however, we will see next that the one proposed here implies that the total current $I^\alpha(t)$ is directly related to the acceleration of the pointer.

Using Eq. (\ref{o.meq1}) and Eq. (\ref{o.meq5}), we can describe the many-particle Schr\"{o}dinger equation of the $M_T$ electrons interacting with the $K$ pointer particles, each one of mass $m$. In order to simplify the notation, we focus on the 1D center of mass of the pointer $\xi$ whose mass is $M=K\; m$. Then, we can write:
\begin{eqnarray}
 i \hbar \frac{\partial \Phi(\vec r,\xi,t)}{\partial t}=
 \Big(-\sum_{k = 1}^{M_T} \frac{\hbar^2}{2m^*}\frac {\partial^2} {\partial \vec r_k^2}  + U(\vec r,t) - \frac{\hbar^2}{2 M}\frac {\partial^2} {\partial \xi^2}  \nonumber \\
+  i \hbar \lambda \frac {Q(\vec r)} {q} \frac {\partial} {\partial \xi}\Big) \Phi(\vec r,\xi,t). \; \; \;
\label{o.meq8}
\end{eqnarray}
At this point, it is relevant to compute the Bohmian velocity of the center of masses. By inserting the polar form of the wave function $\Phi(\vec r,\xi,t) = R(\vec r,\xi,t)e^{\frac{i}{\hbar}S(\vec r,\xi,t)}$ into Eq. (\ref{o.meq8}), we derive the corresponding Hamilton-Jacobi and the continuity equations from which we can define the pointer velocity as
\begin{eqnarray}
v_{\xi}(\vec r,\xi,t) = \frac{1}{M} \frac{\partial S(\vec r,\xi,t)}{\partial \xi} - \lambda \frac {Q(\vec r)} {q}.
\label{o.meq10}
\end{eqnarray}
Since the mass $M$ is very large, the first term in the right hand side of Eq. (\ref{o.meq10}) can be neglected. Therefore, for a particular trajectory $\{\vec r^\alpha(t),\xi^\alpha(t)\}$, it turns out that the acceleration of the pointer,  i.e. the time derivative of Eq. (\ref{o.meq10}), is proportional to the total current of the system defined in Eq. (\ref{o.meq2}):
\begin{eqnarray}
\frac {d v_{\xi}(\vec r^\alpha(t),\xi^\alpha(t),t) } {dt} \approx -\lambda \frac {d  Q(\vec r^\alpha(t))/q } {dt} = -\frac {\lambda} {q} I^\alpha(t).
\label{o.meq11}
\end{eqnarray}

\subsubsection{An example: Numerical solution of system plus apparatus Schr\"{o}dinger equation}
\label{numerical_measure}

\begin{figure}
\centering
\includegraphics[width=0.57\columnwidth]{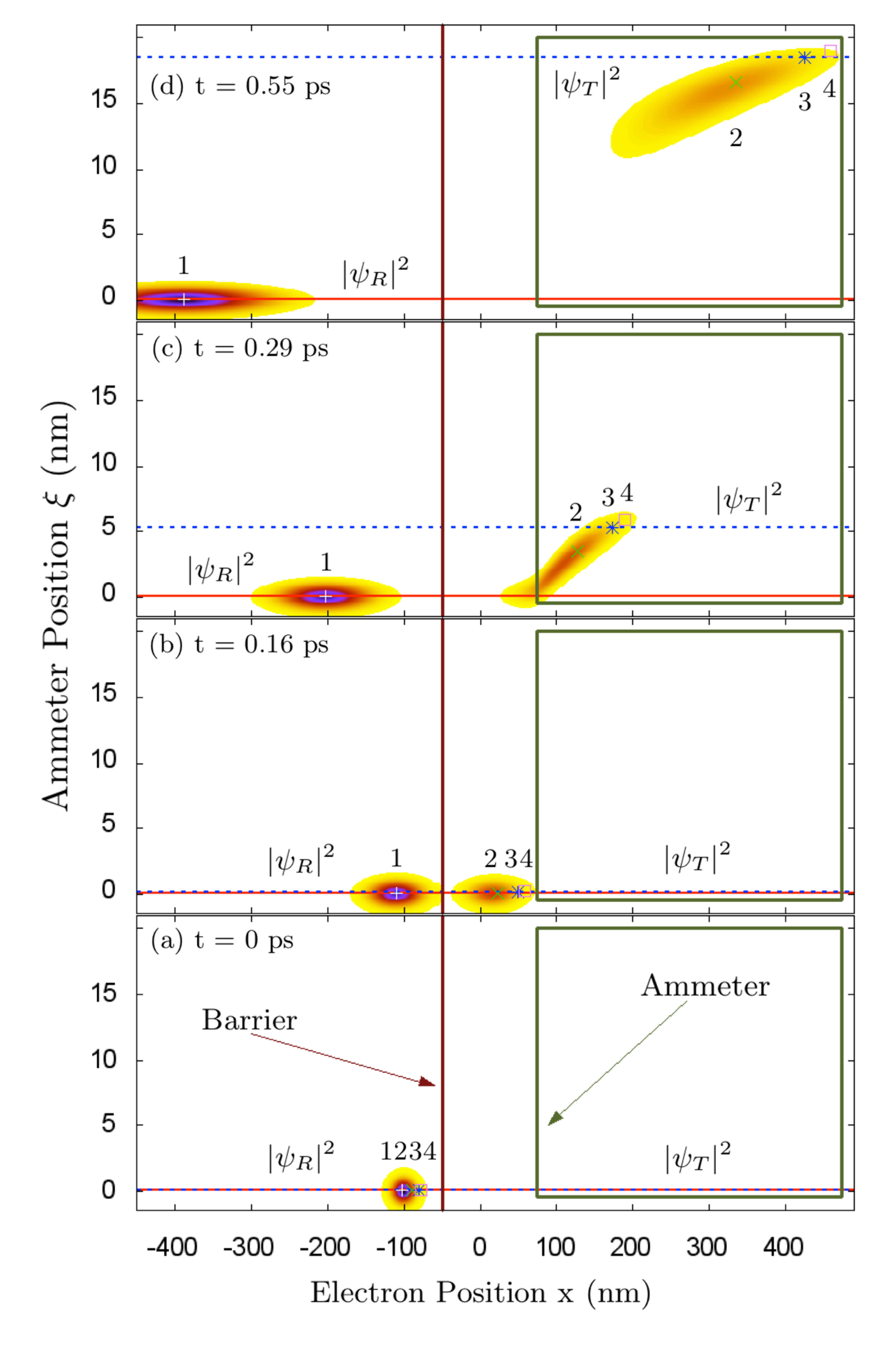}
\caption{(Color online) Time evolution of the squared modulus of $\Phi(x,\xi,t)$ at four different times. The system barrier is indicated by a solid line and the region in the configuration space where the system-apparatus interaction is nonzero by a rectangle. The solid horizontal line indicates the modulus of the (reflected) conditional wave function $|\psi_R|^2=|\Phi(x,\xi^{\alpha=1}(t),t)|^2$, while the dashed horizontal line corresponds to $|\psi_T|^2=|\Phi(x,\xi^{\alpha=3}(t),t)|^2$. The time evolution of four trajectories $\{x^\alpha(t),\xi^\alpha(t)\}$ with different initial positions are presented with with $\square$, $*$, $\times$ and $+$. The transmitted trajectories ($\square$, $*$, and $\times$) at (c) and (d) have different pointer positions associated to $\xi^\alpha(t)$ with $\alpha=2,3,4$ because their evolution does only depend on the transmitted wave packet. The pointer position associated with the reflected trajectory ($+$) with $\alpha=1$ does not move because there is no interaction between this trajectory and the apparatus.}
\label{o.figure5}
\end{figure}

We solve here the 2D version of Eq. (\ref{o.meq8}), where the system and the pointer are described by  just one particle $x$ and $\xi$. Consider that the initial wave function is a product of two Gaussian wave packets. 
The central kinetic energies, central positions and spatial dispersions being respectively $E_x = 0.1 \; eV$, $x_c = -100\; nm$ and $\sigma_x = 8 \; nm$ for the particle, and $E_\xi = 0\; eV$, $\xi_c = 0\; nm$ and $\sigma_\xi = 0.5\; nm$ 
for the pointer. The system consists of an electron (with $m^{*}$ equal to $0.068$ the electron free mass) impinging upon an eckart barrier $U(x,t)=V_0/ \cosh^2[(x-x_{bar})/W]$ with $V_0 = 0.3\; eV$, $x_{bar} = -50\; nm$ and $W = 1\; nm$ 
(see the line at $x=-50\; nm$ in \fref{o.figure5}). The pointer of the apparatus $\xi$ (with $M \approx 75000\; m^*$) interacts with the system through the term $i \hbar \lambda (Q(x)/q) {\partial}/{\partial \xi}$ with $\lambda =50 \; nm/ps$. 
We consider a lateral surface of $S_D= 900\; nm^2$ located at $x_D= 75\; nm$ so that $Q(x)$, defined in Eqs. (\ref{o.meq2}), (\ref{o.meq3}) and (\ref{o.meq4}), is only different from zero on the right hand side of the plots in \fref{o.figure5}. 
We indicate this region by a (ammeter) rectangle in the configuration space.

The numerical solution of the modulus of $\Phi(x,\xi,t)$ is plotted at four different times. At the initial time the entire wave function is at the left of the barrier. Later the wave function is splitted up into reflected and transmitted parts due to the barrier. In Figs. \ref{o.figure5} (a) and (b) the velocity in the $\xi$ direction remains zero because the wave function has not yet arrived to the apparatus. In Figs. \ref{o.figure5} (c) and (d) the interaction of the apparatus with the transmitted part of the wave function shows up. The \emph{local} velocity of the wave packet in $\xi$ rises according to Eq. (\ref{o.meq10}).  In \fref{o.figure5} we also plot the positions of the system and pointer $\{x(t),\xi(t)\}$ for four different trajectories. Each trajectory corresponds to a different experiment. There are three transmitted particles (in the $x$-direction). While the pointer (i.e. $\xi(t)$) does not move for the reflected particles, its evolution  for the transmitted ones clearly shows the correlation with the electrical current of the system.

At this point it would be enlightening to compare the schematic results of \fref{o.figure1}(e) and (f) with the numerical results of \fref{o.figure5}. The transmitted trajectory in \fref{o.figure1}(e) corresponds to any of the particles $\alpha=2$, $\alpha=3$ and $\alpha=4$ of \fref{o.figure5}. Let us take, for example, $\alpha=3$, which is drawn in an horizontal dashed line in \fref{o.figure5}. The wave packet $\psi_T$ shown in \fref{o.figure1}(e) corresponds then to the conditional wave function $\Phi(x,\xi^{\alpha=3}(t),t)$. Following the conditional wave function  $\psi_T=\Phi(x,\xi^{\alpha=3}(t),t)$ in \fref{o.figure5}, we clearly see how the reflected part of the wave packet disappears in the (conditional) configuration space of the system $\{x,\xi^{\alpha=3}(t)\}$. The disappearance of part of the wave packet (collapse) due to the measurement of the electrical current is totally demystified in this (system-apparatus) picture. For the reflected particle in \fref{o.figure1}(d), we follow identical arguments using the $\alpha=1$ particle in \fref{o.figure5}. See $\psi_R=\Phi(x,\xi^{\alpha=1}(t),t)$ in the horizontal solid line of \fref{o.figure5}. 

\begin{figure}
\centering
\includegraphics[width=0.57\columnwidth]{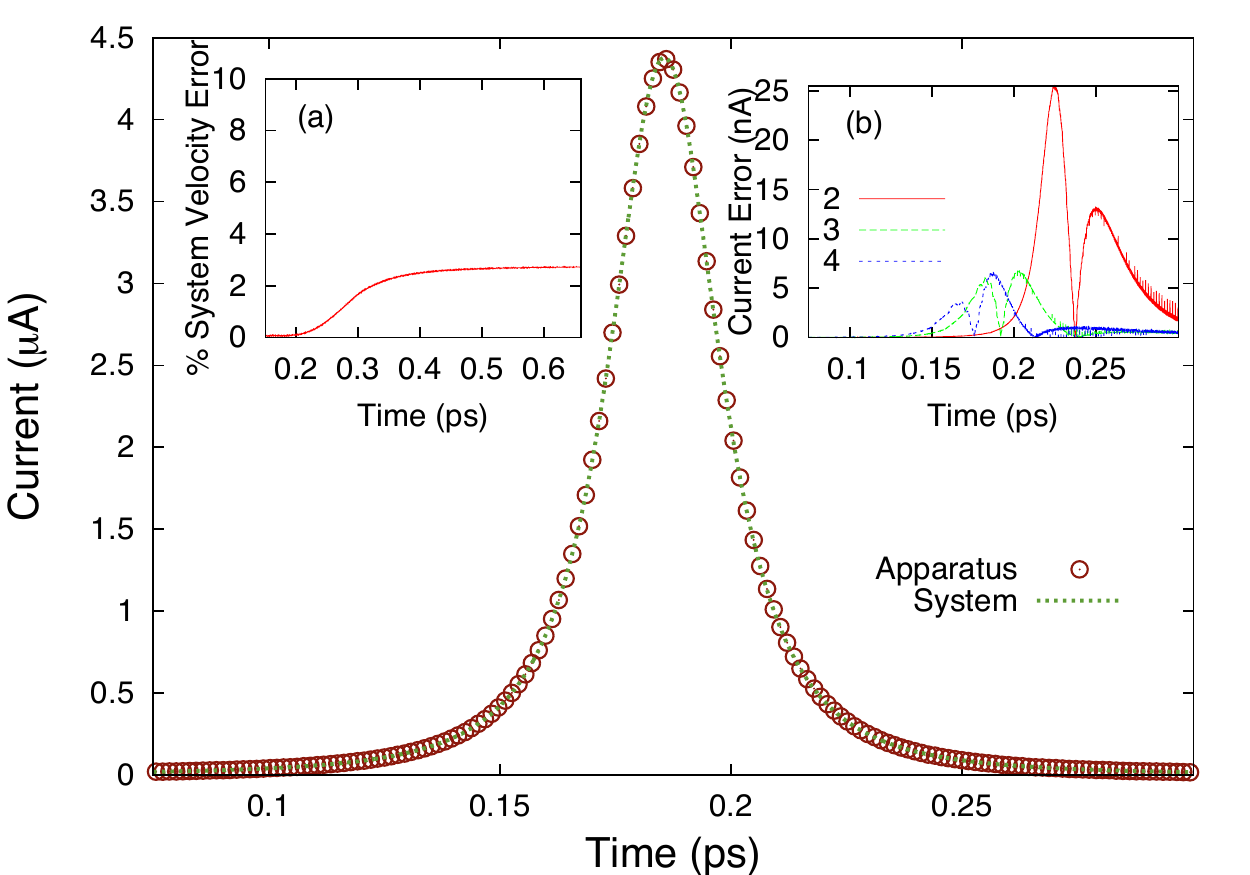}
\caption{(Color online) Total current associated to the $*$ trajectory $\{x^3(t),\xi^3(t)\}$ of \fref{o.figure5} as a function of time computed from the acceleration of the pointer trajectory computed from Eq. (\ref{o.meq8}) in (red) circles and from the system trajectory computed from Eq. (\ref{Scho}) in dotted (green) line. For commodity, we reverse the sign of the current.  (a) Relative (ensemble) error for the system Bohmian velocity when computed from Eq. (\ref{Scho}). (b) Absolute error of the system current of the $\alpha=2$, $\alpha=3$ and $\alpha=4$ trajectories of \fref{o.figure5} when comparing the solution from Eq. (\ref{o.meq8}) (with ammeter) and from Eq. (\ref{Scho}) (without ammeter). }
\label{o.figure6}
\end{figure}

Finally, in \fref{o.figure6}, we plot the \lq\lq{}measured\rq\rq{} current in circles computed from the acceleration of the pointer (as indicated in Eq. (\ref{o.meq10})) for the third trajectory marked with $*$ in \fref{o.figure5}.
As seen in Eq. (\ref{o.meq11}) and \fref{o.figure6}, the evolution of the pointer $\xi^ {\alpha=3}(t)$ describes the evolution of the total current $I^{\alpha=3}(t)$, however, the movement of the pointer (see \fref{o.figure6}) is still not macroscopic (it only moves a few nanometers). It can be easily demonstrated that for other parameters $\lambda$ and $S_D$, $\xi(t)$ would have a macroscopic movement.
It is important to underline that for a realistic ammeter one could expect other features not included in the simple model just presented. Let us discuss this point in detail. As a matter of fact one can expect that the wave function of the system-apparatus will channel into a set of non-overlapping parts, each corresponding to a particular well-defined position of the pointer of the ammeter. As shown in Fig. \ref{o.figure5}, our model, splits the wave function, in the $\xi$ direction, only into two non-overlapping channels. Although not complete, this feature is enough to calculate DC currents and low frequency noise. If one is interested in going beyond this simple model and studying high frequency regimes, other routes can be followed. 
For example, it can be directly considered the many particle Coulomb interaction between the electrons in the active region of the device and the electrons in the cables, this procedure leads to a weak measurement of the total 
current \cite{IWCE,PRA,FNL2016}. 
In any case, the model presented in this section gives enough information for our present purposes: it provides two channels (one for the reflected and one for the transmitted part of the system wave function) and a strict relation between the system and the ammeter (see Eq. \eref{o.meq10}), from this we can infer the perturbation induced by the ammeter on the measured system.

Once the first technical difficulty (i.e. specifying how the ammeter is included in the Hamiltonian) is solved, we must discuss about the difficulties of solving the Schr\"odinger equation including the system and apparatus degrees of freedom. \emph{Can we avoid the inclusion of the pointer degrees of freedom in the Schr\"{o}dinger equation without loosing much accuracy?} The answer to this question is affirmative whenever the apparatus induces a small distortion on the system. The distortion on the system's trajectories can be quantified by defining the relative (ensemble) error of the system velocity:
\begin{eqnarray}
\label{o.meq12}
{Error}(t) = \lim_{M_{\alpha} \to \infty} \frac{\sum\limits_{\alpha=1}^{M_{\alpha}} |v^{\alpha}_{\lambda}(t)-v^{\alpha}_{0}(t)|}{\sum\limits_{\alpha=1}^{M_{\alpha}}|v^{\alpha}_{\lambda}(t)|}. 
\end{eqnarray}
We define $v^\alpha_{\lambda}(t)$ as a Bohmian velocity of the system when we use Eq. (\ref{o.meq8}) with $\lambda=50\; nm/ps$, while $v^\alpha_{0}(t)$ when we do not consider the apparatus ($\lambda=0$) in \eref{o.meq8}. We see in the inset (a) of \fref{o.figure6} that the relative error on the velocity defined in \eref{o.meq12} is less than 3 \%. Then, if we avoid the inclusion of the pointer, we can compute the total current directly from the system trajectory $x^\alpha_{0}(t)$ without apparatus, with a small error. This result is confirmed by the dashed line in \fref{o.figure6} that corresponds to the total current computed from  \eref{o.meq2} using the system trajectories in \eref{o.meq8} when  $\lambda=0$. See the absolute error in the inset (b) of \fref{o.figure6} defined as the difference (in absolute value) between the exact value in circles and the approximate value in a dotted line. The reason why the agreement between the current computed from the pointer trajectory and the system trajectory is so good is because the main distortion of the system trajectory comes from the barrier, not by the apparatus. The former splits the initial wave packet into two separated parts (transmitted and reflected components), while the latter provides a small \emph{adiabatic} perturbation on the system trajectories as seen in \fref{o.figure5}. 
While the error is very small for the simple model pointer used in \fref{o.figure5}, a larger error can be expected in real ammeters. In any case, it perfectly clarifies that we can use the Bohmian trajectories of the system alone (without the pointer) to compute the current when  the back action of the apparatus on the system trajectories is not much relevant.  It is very important to emphasize, however, that the change from Eq. (\ref{o.meq8}) to Eq. (\ref{Scho}) is only technical, without any fundamental implication. 
In summary, when the measurement apparatus has a small effect on the system (e.g. for the computation of DC and low frequency noise) and the pointer position does perfectly specify the value of the total current of the system (as for our model system in \fref{o.figure6}), then we can avoid the explicit simulation of the pointer in order to surpass computational burdens.

\subsection{Coulomb correlations beyond mean field}
\label{Bitlles2}

In general, the Coulomb interaction introduces screening among electrons implying that the total charge in the whole circuit is zero, this is what we call \emph{overall charge neutrality}. Moreover, the proper modeling of the total current has to take into account the dynamics of electrons in a self-consistent way to preserve the \emph{conservation of the total current}. Both conditions, require an (approximate) solution to the many-body Coulomb interaction. Below, we describe how using a small simulation box (including only the active region of the electronic device) both conditions can be preserved \cite{o.albareda2009prb,o.albareda2010prb,JCEhender,JCEL_BCs}.

\subsubsection{The many-particle Poisson equation in the active region}
\label{Poisson_description}

The evaluation of the each term of $U_{a}(x_{a},\vec{r}^\alpha_{b}[t],t)$ in Eq. (\ref{Conditional_eq}) can be written as:
\begin{eqnarray}
U_{a}(\vec r_{a},\vec{r}^\alpha_{b}(t),t) &=&  \sum^{N(t)}_{ j \neq a}\frac{q^{2}}{4 \pi \varepsilon |\vec r_{a}-\vec r^\alpha_{j}(t)|} \nonumber \\
&+&\sum^{M_T}_{j=N(t)+1}\frac{q^{2}}{4 \pi \varepsilon |\vec r_{a}-\vec r^\alpha_{j}(t)|},
\label{exact_coulomb}
\end{eqnarray}
where the first ensemble $\{1,\ldots,N(t)\}$ corresponds to the electrons in the device active region and the second ensemble $\{N(t)+1,\ldots,M_T\}$ corresponds to those that are in the leads and reservoirs. We do not want to deal with the dynamics of the second ensemble. Thus, due to the uniqueness (electrostatic) theorem \cite{o.jackson1962book}, the scalar potential inside the active region can be introduced through the solution of a Poisson equation with a proper definition of its boundary condition. Then, instead of using (\ref{exact_coulomb}), we compute the following 3D Poisson equation inside the active region:
\begin{equation}
\nabla^{2}_{\vec{r}_{a}}\left(\varepsilon(\vec{r_{a}})U_{a}(\vec{r_{a}},\vec{r}^\alpha_{b}(t),t)\right)=\rho_{a}(\vec{r}_{a},\vec{r}^\alpha_{b}(t),t),
\label{Poisson}
\end{equation}
where the charge density $\rho_{a}(\vec{r}_{a},\vec{r}^\alpha_{b}(t),t)$ can be written as:
\begin{equation}
\rho_{a}(\vec{r}_{a},\vec{r}^\alpha_{b}(t),t)=\sum^{N(t)}_{\shortstack{$\scriptstyle j=1$\\
$\scriptstyle j \neq a$}} -q \delta(\vec{r}_{a}-\vec{r}^\alpha_{j}(t)).
\label{density}
\end{equation}
Each $a$-electron is associated to a different Poisson Eq. (\ref{Poisson}) with a different charge density (\ref{density}). See Ref. \cite{o.albareda2009prb} for a detailed discussion on this point. Now, the role of the second term of the right hand side of (\ref{exact_coulomb}), i.e. the interaction of the first ensemble $\{1,\ldots,N(t)\}$ of electrons with the second one $\{N(t) + 1,\ldots,M_T\}$ can be, in principle, represented by a proper set of conditions at the borders of the active region \cite{o.albareda2010prb}. The achievement of overall-charge neutrality will ultimately depend on our ability to properly define these boundary conditions.

\subsubsection{Time-dependent Boundary Conditions of the Poisson equation}
\label{boundaries}
Because the far from equilibrium conditions governing the dynamics of an electronic device, it is very difficult to anticipate an educated guess for the scalar potential (or the electric field) and the charge density on the boundaries of the active region when the leads and the reservoirs are not simulated. Our strategy consists on deriving analytical expressions for all these quantities along the leads and reservoirs to transfer the specification of the boundary conditions at the borders of the active region into a much simpler ones deep inside the reservoirs \cite{o.albareda2010prb,JCEL_BCs}.

In order to derive a set of analytical equations, we only need to take into account two general considerations: first, the total charge in a large volume including the device active region, the leads and the reservoirs, tends to zero within the dielectric relaxation time, $\tau _c  = \varepsilon /\sigma$. 
Then, it can be easily demonstrated that the electric field deep inside the reservoirs, $E_{S/D}^C (t)$, tends to its drift value $E_{S/D}^C (t) \to E_{S/D}^{{\rm drift}} (t)$ within the same time $\tau _c  = \varepsilon /\sigma$. Second, the scalar potentials deep inside the reservoir are fixed by the external bias $V_S^C (t) = 0$ and $V_D^C (t) = V_{{\rm external}} (t)$. With these considerations in mind, the analytical (temporal and spatial) relations for the charge density, the electric field and the scalar potential must be then coupled to an injection model controlling the amount of charge on the borders of the active region. Such a coupled procedure results in a boundary conditions' algorithm for the Poisson equation in (\ref{Poisson}) that enforces overall-charge neutrality and current conservation \cite{o.albareda2010prb,JCEL_BCs}.
\begin{figure}
\centering
\includegraphics[width=0.57\columnwidth]{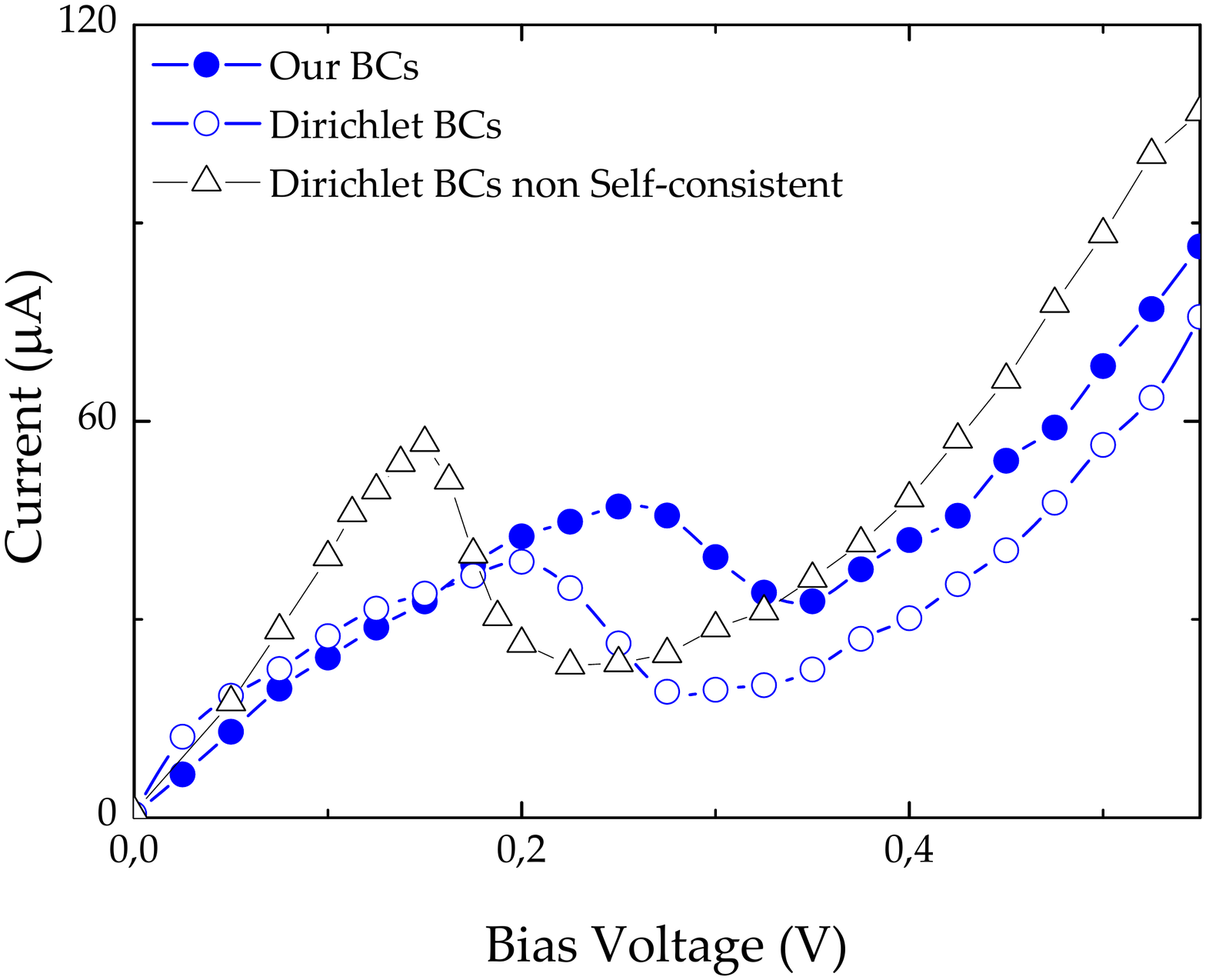}
\caption{(Color online) RTD current-voltage characteristic. Results taking into account the Coulomb correlations between the leads and the device active region are presented in solid circles. Open circles refer to the same results neglecting the lead-active region interaction. Open triangles refer to a single-particle scenario. Transport takes place from source to drain in the $x$ direction with lateral dimensions $L_{y}=L_{z}=48.6nm$. Room temperature is assumed.}
\label{o.figure7}
\end{figure}

In order to highlight the importance of the Coulomb interaction in the prediction of the current-voltage characteristic of nanoelectronic devices, let us apply the above concepts to simulate a simple Resonant Tunneling Device (RTD) \cite{o.albareda2010prb}. We consider two highly doped drain-source $GaAs$ regions (the leads), two $AlGaAs$ barriers, and a quantum well (the device active region). All Schr\"{o}dinger equations (\ref{Conditional_eq}) are coupled to its own Poisson equations (\ref{Poisson}) with the boundary conditions given above. How the electrical current is computed will be explained in detail in \sref{Pract_comp}, here we will directly focus on the results. We distinguish between (i) the Coulomb interaction among electrons inside the active region and (ii) the Coulomb correlations among these electrons and those outside the active region. In \fref{o.figure7}, we compare the characteristic DC obtained by including all Coulomb correlations with those obtained, first, by neglecting the dynamical Coulomb correlations between electrons inside and outside the active region (i.e. assuming standard Dirichlet boundary conditions), and second, switching off all Coulomb correlations (i.e. single-particle treatment of electron transport). The differences between the three approaches appear not only in the magnitude of the current but also in the position of the resonant region. More details can be found in Ref. \cite{o.albareda2010prb}.

\subsection{Stochastic injection of electrons}
\label{Bitlles3}

Since a large part of the degrees of freedom of a circuit shown in \fref{o.figure4} is neglected (we want to explicitely simulate only the active region), we cannot completely specify the initial $N(t)$-particle wave function inside the active region (we do not know with certainty the number of electrons $N(t)$, their energies, etc.). We can assume, however, that the mean energy of injected electrons follows Fermi-Dirac statistics. We thus include the probabilities of these states by introducing an additional probability distribution $h$. From a practical point of view, this means that apart from the uncertainty in the initial position of the quantum trajectories (the $\alpha$-distribution mentioned in \sref{Intro}), we do also have an additional uncertainty on the properties of the injected electrons, $h = \{1, ..., M_h\}$, that arises because the active region is an open system. Strictly speaking, because of the $h$ distribution, we are no longer dealing with a single pure $N(t)$-particle state, but with a mixed quantum system prepared by statistically combining different pure states of $N(t)$ particles. The microscopic description of the system, however, implies that we have a perfect knowledge of each of the $N(t)$ particle states and that the statistical ensemble of the density matrix is explicitly taken into account.

\begin{figure}
\centering
\includegraphics[width=0.57\columnwidth]{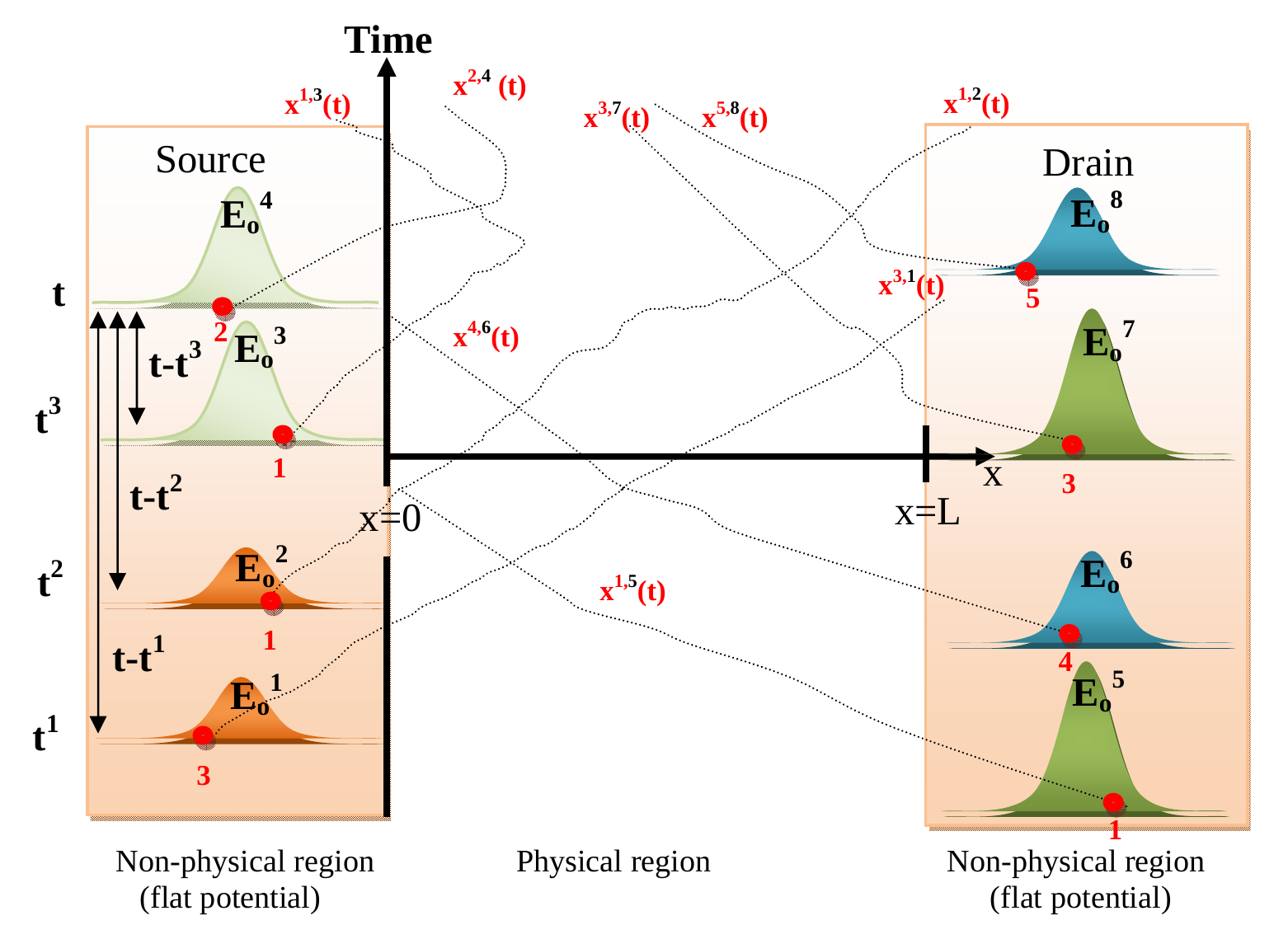}
\caption{(Color online) Schematic representation of the (physical $0<x<L$ and unphysical) simulation box for a particular selection of the parameters $E_o^h$ and entering times $t^h$ of the initial Gaussian wave packets  ($h$-distribution)  and the initial positions $x^{\alpha,h}(0)$ of the Bohmian trajectories ($\alpha$-distribution). The wave packets $h=1$ and $h=2$ have identical energies $E_o^1=E_o^2$ but different injecting times $t^1 \neq t^2$. The injection model does also assume that the exact $h-$distribution (except for an irrelevant time offset in $t^h$) plotted here will be repeated in a later time, but with different $x^{\alpha,h}(0)$, accomplishing the quantum equilibrium of the $\alpha-$ distribution.}
\label{o.figure8}
\end{figure}

Our Bohmian protocol requires each electron to be described by a (conditional) wave function ($h$ distribution) plus a Bohmian trajectory ($\alpha$ distribution). Every time an electron with a particular initial wave function is selected to enter the device active region according to the $h$ distribution, an initial position $x^{\alpha,h}(0)$ associated to this ($h$ selected)  wave packet has to be randomly selected according to \eref{Equil_hypo}. See Appendix \ref{appendixA} for numerical details of the wave packet and trajectories computation. Next, we explain how the $h$ distribution of the wave packets is defined. For each contact, we select a flat potential region in an (non-physical) extension  (for $x<0$ in the source and $x>L$ in the drain) of the simulation box (see \fref{o.figure8}). The initial Gaussian wave packet\footnote{In a flat potential region without interaction, the single-particle wave packets are exactly the normalized conditional (Bohmian) wave functions discussed in this work.} is defined in this flat potential region (deep inside the contact). In particular, the central position of all wave packets is selected $x_c=100\;nm$ far from the border of the active region and the spatial dispersion of the wave packet is $\sigma_x=25\; nm$ (i.e. the wave packet is somehow similar to a scattering state). The only two additional $h$-parameters that we still have to fix to fully define the wave packet are the central kinetic energy $E_o$ of the wave packet and the injecting time $t_o$ when the electron \emph{effectively} enters the simulation box. The selection of the energy $E_o$ has to satisfy the Fermi-Dirac occupation function $f (E_o)$ that depends on the (quasi) Fermi energy and temperature. The selection of the time when the electron is injected is a bit more complex \cite{o.oriols2007see}.

In order to simplify our explanation, let us assume a 1D system with parabolic bands where the (central) kinetic energy $E_o$ of the wave packet is related to the (central) wave vector $k_{o}$ by $E_o=(\hbar k_{o})^2/(2 m^*)$.  Let us define $t_o$ as the minimum temporal separation between the injection of two wave packets whose central wave vectors and central positions fit into the following particular phase-space cell $k_o  \in [k_b ,\,\,k_b  + \Delta k )$ and $x_c \in [x_b ,\,x_b  + \Delta x)$, $x_b$ being the border of the simulation region. For a 1D system, the value of $t_o$ can be easily estimated. The number of electrons $n_{1D}$ in the particular phase space cell $\Delta k \cdot\Delta x$ is $n_{1D}  = 2\cdot\Delta k \cdot\Delta x/(2\pi )$ where the factor  2 takes into account the spin degeneracy. These electrons have been injected into $\Delta x$ during the time interval $\Delta t$ defined as the time needed for electrons with velocity $v_x  = \Delta x/\Delta t = \hbar \,k_o /m^*$ to travel a distance $\Delta x$. Therefore, the minimum temporal separation, $t_o$, between the injection of two electrons into the previous cell is $\Delta t$ divided by the maximum number $n_{1D}$ of electrons:
\begin{equation}
\label{eq::Fermi2}
 {t_o}  = \frac{{\Delta t}}{{n_{{\rm{1D}}} }} = \left( {\frac{1}{\pi }\frac{{\hbar \,k_o }}{{m^* }}\,\,\Delta k } \right)^{ - 1}.
\end{equation}
It is very instructive to understand the minimum temporal separation $t_o$ in (\ref{eq::Fermi2}) as a consequence of the wave packet version of the Pauli principle.  The simultaneous injection of two electrons with similar central positions and central momentums would require such a huge amount of energy that its probability is almost zero. In other words, a subsequent electrons with central positions and central momentums equal to the preceding ones can only be injected after a time interval given by $t_o$. This is the time interval necessary to ensure that the first electron has traveled a distance $t_o  \times \hbar \; k_o /m^* $ so that the second one is located in a different central position (no Pauli principle). The injection of electrons (from the mentioned phase-space cell) at multiple times of $t_o$ depends finally on the statistics imposed by the Fermi-Dirac function mentioned above. During each attempt of injection at multiples of $t_o$, we select a random number $r$, and the electron is effectively injected only if $f(E_o)>r$. The mathematical definition of the  rate and randomness of the injection process are given by the following binomial probability $P(E_o,N,\tau)$ \cite{o.oriols2007see}:
\begin{equation}\label{eq::Fermi1}
P(E_o,N,\tau ) = \frac{{M_\tau  !}}{{N!\cdot(M_\tau   - N)!}}f(E_o)^N \left( {1 - f(E_o)} \right)^{M_\tau   - N}.
\end{equation}
This expression defines the probability that N electrons (from the mentioned phase-space cell) are \emph{effectively} injected into the active region during the time interval $\tau$. The parameter $M_\tau$ is the number of attempts of injecting electrons during this time interval $\tau$, defined as a natural number that rounds the quotient $\tau /t_o$ to the nearest natural number towards zero. The number of injected electrons can be $N = 1,2,.... \le M_\tau$.  In order to satisfy the requirements of \sref{boundaries} the final injection model can be a little more elaborated. More details can be found in Ref. \cite{o.albareda2010prb}.

According to \fref{o.figure8}, we can understand the contacts as large reservoirs of electrons that are waiting to enter the active region.  
The parameters $x_c$, $\sigma_x$, $E_o$ (or central wave vector) and a time $t^h$, defined as the entering times,  are assigned to each initial wave packet. Additionally, we give an initial position $x^{\alpha,h}(0)$ to each electron. The measured value of the current related with the Bohmian trajectories is thus ultimately associated to two sources of uncertainty, $\alpha$ and $h$, i.e. $I^{\alpha,h}(t)$. In order to give a precise recipe to compute the electrical current and its moments, from now on we explicitly write the whole indexation.

\section{Computation of the Electrical Current and its Moments with BITLLES} \label{Pract_comp}
\label{Computation}

This section is devoted to provide a practical method to evaluate the electrical current with the BITLLES simulator. Let us recall here that any kind of information related with the electron transport taking place in the electronic device (such as power consumption, kinetic and potential distributions, etc.) can be easily computed with BITLLES. After a brief dissertation on the limitations of the information contained in the values of the DC currents in quantum systems and the importance of multi-time measurements to predict its fluctuations, in \sref{recipe} we will provide a detailed description of the equations required to evaluate, DC, AC, transients, and current fluctuations. Let us recall here that the mathematical formalism introduced in this section does not differ much (in form) from the semiclassical one. The reason is simple, both descriptions use ensembles of trajectories to compute observables. It is in this regard that the BITLLES simulator includes also a semiclassical Monte Carlo package to simulate electronic devices by solving the Boltzmann transport equation \cite{JAP,intech,JSTAT,APL}.

\subsection{Preliminar considerations}
\label{preliminar}

The prediction of the DC current measured in a laboratory can be computed using two different protocols \cite{o.ferry2013jecl}. First, we can compute $I_{DC}$ by time-averaging the measured value of the total current $I(t)$  from a \emph{unique device} during a large (ideally infinite) period of time $\tau$ :
\begin{equation}
I_{DC}= lim_{\tau \rightarrow  \infty}\frac 1 \tau \int_{0}^{\tau} I(t) dt.
\label{eq2}
\end{equation}
We can use the conduction current  $I_p(t)$  instead of the total current  $I(t)$ in Eq. (\ref{eq2}) because $\frac 1 \tau \int_{0}^{\tau} I_d(t) dt \rightarrow 0$  for large $\tau$  (the displacement current is proportional to a time-derivative of the electric field).

Second\footnote{Strictly speaking, no ergodic theorem exists for an out of equilibrium system \cite{o.price1965book}. Therefore, the ergodic connection between (\ref{eq2}) and (\ref{eq4}) has to be considered as only a very reasonable approximation for DC transport, but not as an exact result.}, we can compute $I_{DC}$  from an ensemble-average of all possible eigenvalues of the current $I_i$ that satisfy the equation $\hat I |\psi_i \rangle =I_i |\psi_i \rangle $ for a particular operator $\hat I$, at one particular time  $t$:
\begin{eqnarray}
I_{DC}= \langle \psi(t) |  \hat I | \psi(t) \rangle = \sum_{j}^M c^*_j(t)  \langle \psi_j | \sum_{i=1}^M I_i · c_i(t) |  \psi_i \rangle  =\sum_{i=1}^M I_i P(I_i),
\label{eq4}
\end{eqnarray}
where we have used the orthonormal property of the eigenstates  $ \langle \psi_j|\psi_i \rangle =\delta_{i,j}$ and the definition of the (Born) probability $P(I_i)=|c_i(t)|^2$ . When the operator $\hat I$  is the conduction current density operator, $\hat I_p=|r \rangle  \langle r|\hat p+\hat p |r \rangle  \langle r|$  with $|r \rangle  \langle r|$  and $\hat p$ the position and momentum operators, then the ensemble-average value  $ \langle \psi(t) | \hat I | \psi(t) \rangle $ gives the well-known result \cite{o.cohen1978book}:
\begin{eqnarray}
I_{DC}= \langle  I_p(t) \rangle =\int_{S_i} \vec J(\vec r,t) d\vec s= \int_{S_i}  \langle \psi(t) | \hat I_p | \psi(t) \rangle  d\vec s \nonumber\\
=\int_{S_i} \frac {\hbar} {m} Im\left(\psi^*(\vec r,t) \vec \nabla \psi(\vec r,t) \right) \; d\vec s. \;\;\;\;\;\;
\label{eq5}
\end{eqnarray}
The quantum expression of the current probability is defined in expression \eref{om.current} of Appendix \ref{appendixC}. Expression (\ref{eq5}) is extraordinarily useful and simple because it allows us to predict the DC of quantum devices without knowing the eigenstates $|\psi_i \rangle $  or eigenvalues  $I_i$ of the current operator  $\hat I_p$. However, the amount of information contained in \eref{eq5} is limited.

\begin{figure}
\centerline{
\includegraphics[width=0.57\columnwidth]{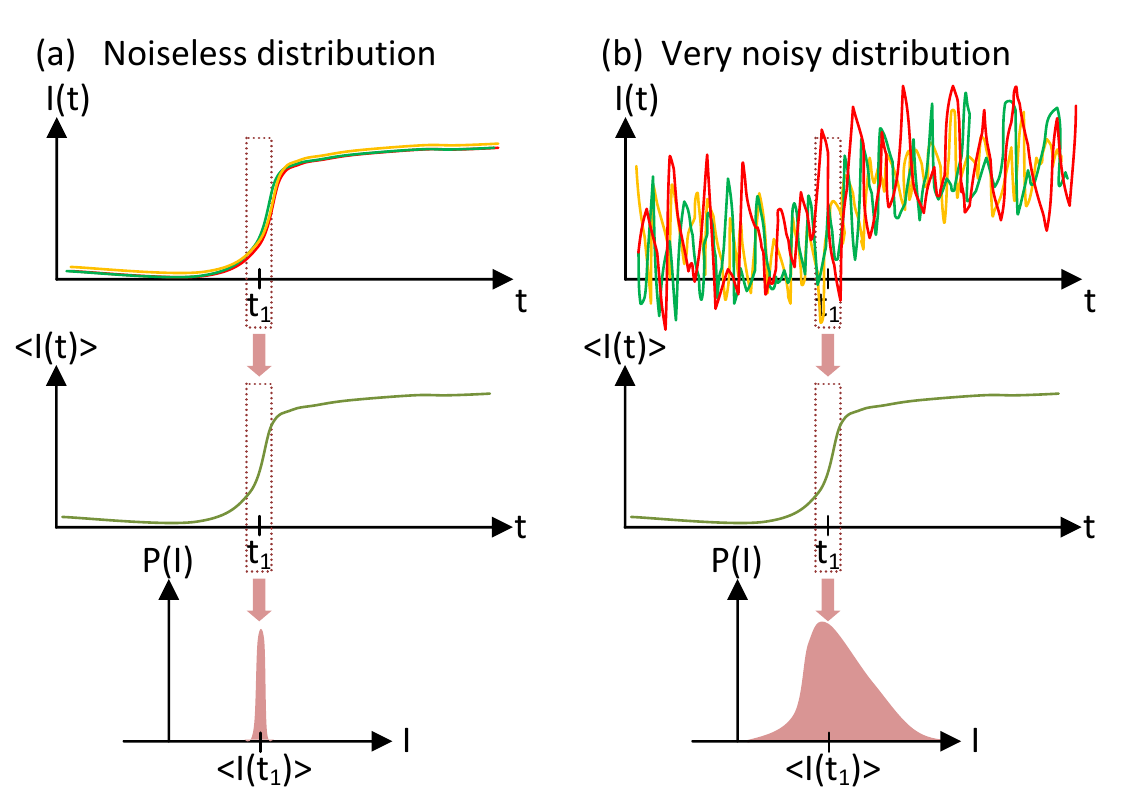}}
\caption{ (Color online) Schematic representation of the measurement of transient currents $I(t)$ on a sample. The ensemble average current $\langle I(t_1) \rangle$ at time $t_1$ can provide misleading information because different current probability distributions provide identical average values. The current probability distribution  P(I) at time $t_1$ provides the complete information of the transport process. However, its computation requires a proper treatment of unitary and non-unitary evolution of the quantum system.}
\label{o.figure9}
\end{figure}

In order to illustrate how reliable is the information provided by expression (\ref{eq5}) to the electronic industry, let us discuss the transient from ``OFF'' to ``ON'' values in the drain-source current of a digital quantum FET when the gate contact voltage suddenly changes from $V_{OFF}$  to  $V_{ON}$. In Fig. \ref{o.figure9}, we plot the measured current $I(t_1)$ and the ensemble current  $ \langle I(t_1) \rangle $. The first problem appears because  $ \langle I(t_1) \rangle$ is not the measured current in a laboratory in a single transistor. The difference between $I(t_1)$ and $ \langle I(t_1) \rangle$ can be irrelevant for a large signal-to-noise ratio (for example, in FETs with many electrons) as seen in Fig. \ref{o.figure9}(a), but it will be certainly meaningful for low signal-to-noise ratios (for example, in FETs with very few electrons) as seen in Fig. \ref{o.figure9}(b). The second problem when using (\ref{eq5}) for AC, transients and noise at high frequencies is that the total time-dependent current cannot be computed only from $\int_{S_i} \vec J(\vec r,t) \; d\vec s$  alone,  but one has to include the displacement current, with all its computational difficulties. The third problem appears because in scenarios where the system is measured many times, we have to take into account the unitary time-evolution determined by the Schr\"{o}dinger equation plus the non-unitary evolution due to the measurement.  We briefly discuss this issue below.

In general, one could argue that the ensemble-average value of such fluctuations can be quantified as:
\begin{eqnarray}
\Delta I=\sum_{i=1}^M (I_i-I_{DC})^2 · P(I_i).
\label{eq6}
\end{eqnarray}
This expression can be related to an ensemble of systems with a unique time measurement of the squared current. Thus, in principle, no multi-time measurement is needed for evaluating (\ref{eq6}). However, all electronic devices can work properly only below a particular cutoff frequency, behaving as a low-pass filter. Therefore, electronic systems are not able to measure \lq\lq{}all\rq\rq{} the noise appearing in (\ref{eq6}), but only the noise whose components have a frequency lower than the measuring device (or the system under test) bandwidth. Therefore, what we can measure in a laboratory is not expression (\ref{eq6}), but the power spectral density of the noise below the mentioned cutoff frequency \cite{o.ferry2013jecl}. Such a power spectral density is related, through a Fourier transform, to the correlation function. The correlation function is the ensemble-value of an event defined as measuring the current  $I_i(t_1)$ at time  $t_1$  and the current $I_j(t_2)$  at time  $t_2$:
\begin{eqnarray}
 \langle I(t_2) I(t_1) \rangle =\sum_{i}^M \sum_{j}^M I_j(t_2)I_i(t_1)  P\left(I_j(t_2),I_i(t_1)\right).
\label{eq7}
\end{eqnarray}
The computation of the probability $ P\left(I_j(t_2),I_i(t_1)\right)$  of this two-time measurement process has to be done carefully. The mandatory perturbation of the state because of the measurement, is the reason why modeling the measurement process plays a fundamental role in determining the noise, even the low (zero) frequency noise. In what follows, we show how the BITLLES simulator computes, in practice, any moment of the total, particle plus displacement, current without the necessity of introducing any kind of non-unitary evolution or using second quantization tools.

\subsection{Practical Method to compute DC, AC, transients and higher moments}
\label{recipe}

As mentioned in \sref{Intro2}, at GHz or THz frequencies, the computation of the current and its fluctuations requires to deal with both the conduction and displacement components. The total current $I(t)$ can be computed from the following two terms:
\begin{eqnarray}
I(t)=\int_{S}  \vec J(\vec r,t) \cdot d\vec s+\int_{S}  \epsilon (\vec r)\frac{\partial \vec E(\vec r,t)}{\partial t} \cdot d\vec s,
\label{displ}
\end{eqnarray}
with $\epsilon (\vec r)$ the (inhomogeneous) electric permittivity.  Alternatively, the Ramo-Shockley-Pellegrini theorem \cite{o.shockley1938jap,o.ramo1939pire,o.pellegrini1986prb,o.pellegrini1993nca,o.pellegrini1993ncb}, briefly described in the Appendix \sref{appendixB}, provides an excellent tool to rewrite expression (\ref{displ}) with an explicit dependence on the electron (Bohmian) velocity $v^k_x(t)$. For a two-terminal device with (ideal) metallic contacts,  the total current generated by $N(t)$ electrons inside the active region of length $L$ at time $t$ can be written as:
\begin{equation}
 I(t)=q \sum_{k=1}^{N(t)} \frac {v^k_x(t)} {L}.
 \label{ivelo}
\end{equation}
Outside the active region, in the metallic contacts, the $k-$electron is immediately thermalized and it does no longer contribute to the current. Expression (\ref{ivelo}) will be used in the rest of the chapter for the computation of the current (its derivation can be found in Appendix \ref{appendixA}).

\subsubsection{Computing DC, AC and transients}
\label{DC}

Taking into account the quantum equilibrium (\ref{Equil_hypo}) and the additional statistical distribution $h$, we can determine the expectation value of the electrical current at time $t_1$ from the following ensemble average:
\begin{equation}
\langle I(t_1) \rangle=\lim_{M_\alpha,M_h\rightarrow\infty} \frac {1} {M_\alpha M_h} \sum_{\alpha=1}^{M_\alpha} \sum_{h=1}^{M_h} I^{\alpha,h}(t_1).
\label{alarcon:Equation6_13}
\end{equation}
Therefore, the procedure to compute the average current would be the following:
\begin{enumerate}
\item At $t=0$, we select a particular realization of the $h$-distribution and a particular realization of the $\alpha$-distribution (see \fref{o.figure8}).
\item We solve the (conditional) Schr\"{o}dinger equation from time $t=0$ till $t=t_1$ (see \sref{CWF}).
\item From (\ref{ivelo}), we compute the value $I^{\alpha,h}(t_1)$.
\item We repeat the steps 1 till 3 for the whole ensemble $\alpha=\{1,...,\infty\}$ and $h=\{1,...,\infty\}$ to evaluate (\ref{alarcon:Equation6_13}).
\end{enumerate}

When the bias is fixed to a constant value, the whole circuit becomes stationary. For a stationary process, the mean current in (\ref{alarcon:Equation6_13}) is independent of time.
Then, if the process is ergodic, we can compute the mean current from the following (first-order) time average expression:
\begin{equation}
\overline{I^{\alpha,h}(t)}=\lim_{T \rightarrow \infty} \frac {1} {T} \int_{-T/2}^{T/2} I^{\alpha,h}(t)dt.
\label{alarcon:Equation6_14}
\end{equation}
In this case, the practical procedure for the computation of the mean current is simpler. Before beginning the simulation, we select only one particular realization of the $h-$distribution for an \textit{infinite}\footnote{The practical procedure for the \textit{infinite} number is selecting a number large enough so that the mean current remains practically unchanged for successive times.}  number of electrons. Simultaneously, we fix the $\alpha-$distribution of the initial positions for the previous (infinite) realization of wave packets.
See \fref{o.figure8} where we have represented a scheme of a single $h-$ and $\alpha-$ element of the distributions in the simulation box in position and time.

A single sample function often provides little information about the statistics of the process. However, if the process is assumed to be ergodic, i.e., time averages equal ensemble averages, then all statistical information can be derived from just one sample element of the $h-$ and $\alpha-$ distributions. To compute transients or AC, the circuit is no longer stationary. Then, ergodicity cannot be assumed and the mean value of the current at each particular time, $t_1$, can be only computed from the ensemble average in (\ref{alarcon:Equation6_13}).

\subsubsection{Computing current fluctuations and higher moments}
\label{fluctuations} 

Let us consider now the problem of providing reliable information to the electronic industry about the switching time needed to differentiate the ``OFF'' and ``ON'' values of the current in a digital quantum FET (drawn schematically in Fig. \ref{o.figure9})\footnote{It should be noted that this switching time is quite different from the delay time discussed in VLSI, and this switching is considerably longer than the delay time.}.  As discussed at the end of Sec. \ref{preliminar}, we feel uncomfortable with providing ensemble-average information about the transient performance of the FET. The reason is because the ensemble-average value $\langle I(t_1) \rangle$  at a particular time  $t_1$ can be misleading. We can get the value  $\langle I(t_1) \rangle$  because the measured currents $I(t_1)$ in all samples are very close to  $\langle I(t_1) \rangle$,  as we plot in Fig. \ref{o.figure9}(a), or because  $I(t_1)$ fluctuates a lot among different values, as plotted in Fig. \ref{o.figure9}(b).  Certainly, it would be mandatory to provide the electronic industry with new complementary information. For example, the statistical variations around the value $\langle I(t_1) \rangle$. This difference is well understood in the classical logic world where one is concerned specifically about noise margins in logic gates. However, the importance of knowing the fluctuations is far less appreciated when dealing with quantum devices.  For example, we can be interested in knowing whether the fluctuations greater than the mean value  $\langle I(t_1) \rangle$ occur more or less often than the lower ones. The skewness is a measure of the asymmetry of the probability distribution of the (random variable) current. See the symmetry of the function $P(I)$ in Fig. \ref{o.figure9}(a) and the asymmetry in Fig. \ref{o.figure9}(b) with positive skew. The quantum prediction of the skewness requires computing correlations among the value of the current measured during three consecutive times.  This means that one must have a good operator representation for these currents at different times, and have confidence that we understand exactly how these operators evolve over the time of interest in the correlation functions.

The consideration of the transport dynamics affected by the measuring process becomes fully pertinent to understand the fluctuations of the current around its DC value. While the \emph{orthodox} explanation of quantum phenomena would have to be dealt with the second law of quantum mechanics (i.e. the collapse of the wave function), our trajectory-based formulation does not require any non-unitary additional evolution because the evolution of the system is considered within an enlarged configuration space that includes the measuring apparatus (see the discussion in \sref{Bohm2}). Therefore, once we know $I^{\alpha,h}(t)$ at any time, the probability of each current value and any higher moment of the current distribution can be straightforwardly computed. This is another very relevant advantage of using quantum trajectories to study quantum electron transport over \textit{orthodox} techniques. The algorithm to compute the current fluctuations is quite simple. The fluctuating signal of the current can be defined from $\Delta I^{\alpha,h}(t)= I^{\alpha,h}(t)- \langle I^{\alpha,h}(t) \rangle$. We can obtain information of the noise from the variance (or the mean square or the second moment) defined as $\langle \Delta I(t)^2 \rangle=\langle I(t)^2 \rangle - \langle I(t) \rangle ^2$. However, experimentalists are interested in having information on how the noise is distributed along the different frequencies\footnote{Most of electronic apparatuses, and the ammeter itself, have to be interpreted as low-pass filters. Therefore, they are not able to measure \textit{all}  noise of the spectrum, but only up to a maximum frequency}. The fluctuations of the current are computed from the covariance:
\begin{eqnarray}
\langle \Delta I(t_1)\Delta I(t_2) \rangle = 
 \lim_{M_\alpha,M_h\rightarrow\infty} \frac {1} {M_\alpha M_h} \sum_{\alpha=1}^{M_\alpha} \sum_{h=1}^{M_h} \Delta I^{\alpha,h}(t_1) \Delta I^{\alpha,h}(t_2).
\label{alarcon:Equation6_16}
\end{eqnarray}
If the process is ergodic, i.e. $\langle \Delta I^{\alpha,h}(t) \Delta I^{\alpha,h}(t+\tau) \rangle =\overline{\Delta I_i(t)\Delta I_i(t+\tau)} $, we can compute the noise equivalently from the autocorrelation function:
\begin{eqnarray}
\overline{\Delta I(t)\Delta I(t+\tau)} =
\lim_{T \rightarrow \infty} \frac {1} {T} \int_{-T/2}^{T/2} \Delta I^{\alpha,h}(t)\Delta I^{\alpha,h}(t+\tau)dt.
\label{alarcon:Equation6_17}
\end{eqnarray}
In addition, a process is called wide-sense (or weakly) stationary if its mean value is constant and its autocorrelation function depends only on $\tau=t_2-t_1$. Then, we define the autocorrelation function $R(\tau)$ as:
 \begin{equation}
 R(\tau)=\overline{\Delta I(t)\Delta I(t+\tau)},
 \label{alarcon:Equation6_17b}
 \end{equation}
because depends only on $\tau=t_2-t_1$.
Wide-sense stationary processes are important because the autocorrelation function in \eref{alarcon:Equation6_17b} and the power spectral density function $S(f)$ (measured by experimentalists) form a Fourier transform pair:
\begin{equation}
S(f)=\int_{-\infty}^\infty R(\tau)e^{-j2 \pi f \tau} \ d\tau.
\label{alarcon:Equation_s(f)}
\end{equation}
This is known as the Wiener-Khinchine theorem. In many systems, one obtains the well known Schottky's result \cite{o.Schottky} or Poissonian shot noise, $S_{shot}(0)=2q\left\langle I\right\rangle$.
\begin{figure}
\centering
\includegraphics[width=0.45\textwidth]{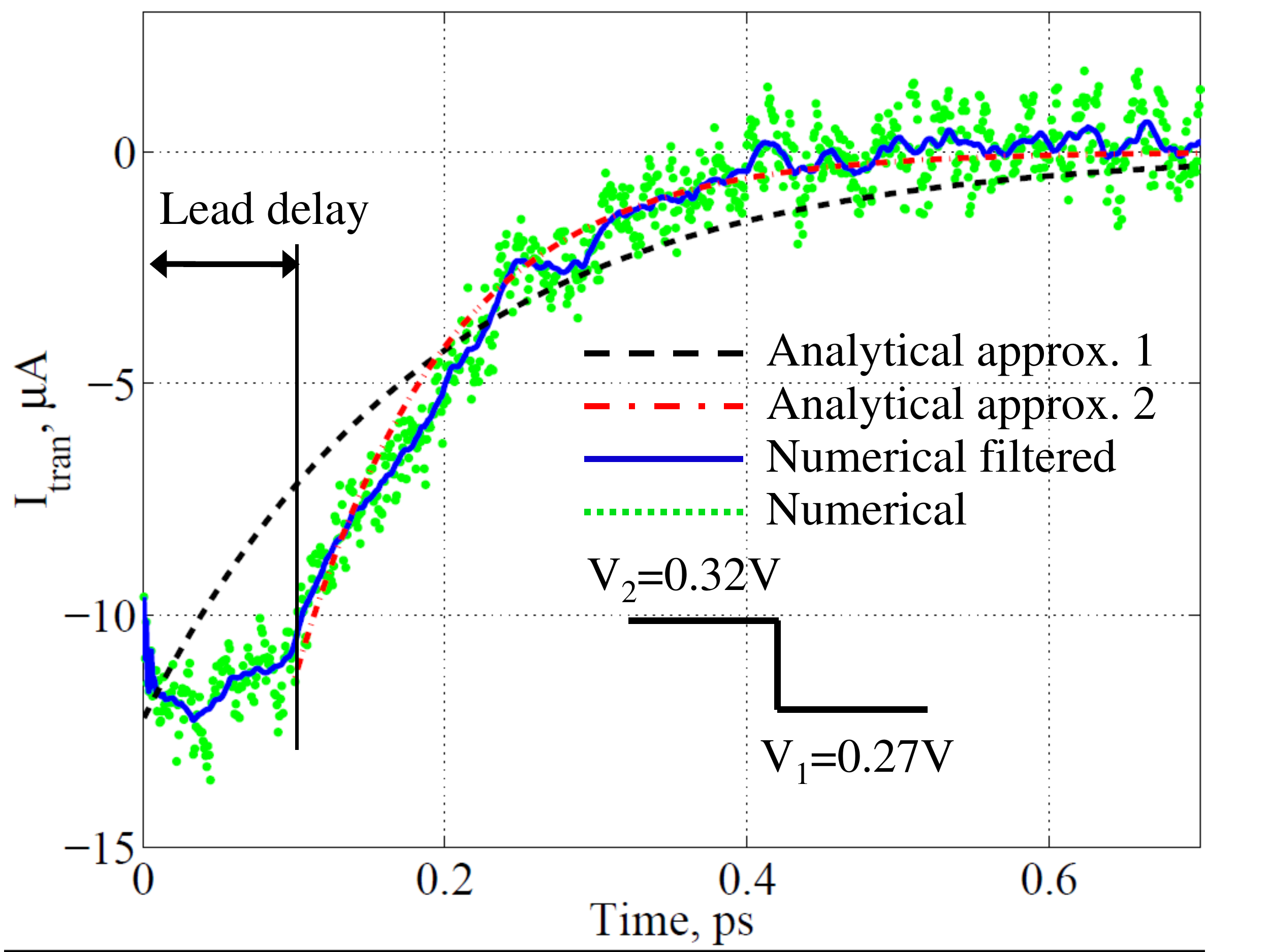}
\caption{ (Color online) Transient current $I_{tran}(t)$ computed analytically and numerically from (\ref{alarcon:Equation6_13}).}
\label{o.figure10}
\end{figure}

\begin{figure}
\centering
\includegraphics[width=0.57\columnwidth]{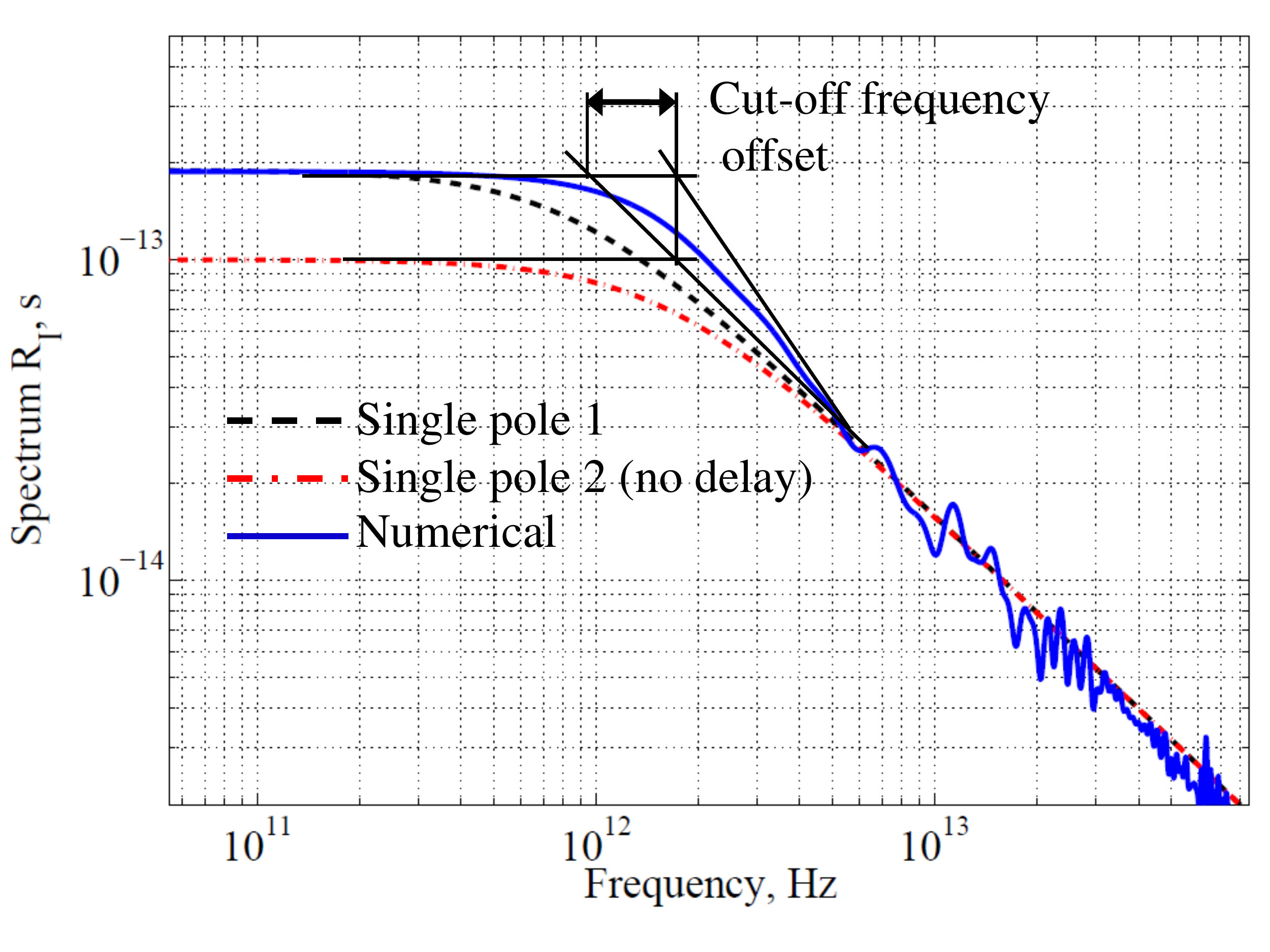}
\caption{ (Color online) Fourier transform of $I_{tran}(t)$ of \fref{o.figure10}. A logarithmic scale is used to resolve the cutoff frequency offset.}
\label{o.figure11}
\end{figure}

As a practical example of the computation of the fluctuations, we show here the current response to a the step input voltage in the Negative Differential Conductance region of a RTD. The input signal is the step voltage $V(t)=V_{1}u(t)+V_{2}\left[  1-u(t)\right]$ where $u(t)$ is the Heaviside (step) function. The voltages $V_{1}$ and $V_{2}$ are constant. Then the current response can be expressed as $I(t)=I_{tran}(t)+I_{1}u(t)+I_{2}\left[  1-u(t)\right]$ where $I_{1}$ and $I_{2}$ are the stationary currents corresponding to $V_{1}$ and $V_{2}$ respectively, and $I_{tran}$ is the \textit{intrinsic} transient current.
The results are reported in \fref{o.figure10} where $I_{tran}(t)$ manifests a delay with respect to the step input voltage, due to the dynamical adjustment of the electric field in the leads. After this delay, the current response becomes a RLC-like response (dot-dashed line) i.e. purely exponential. Performing the Fourier transform of $I_{tran}(t)$ in \fref{o.figure11} and comparing with the single pole spectra (Fourier transform of the RLC-like responses, dashed and dashed dotted lines), we are able to estimate the cutoff frequency and the frequency offset due to the delay \cite{o.fabio2011}.

\begin{figure}
\centering
\includegraphics[width=0.57\columnwidth]{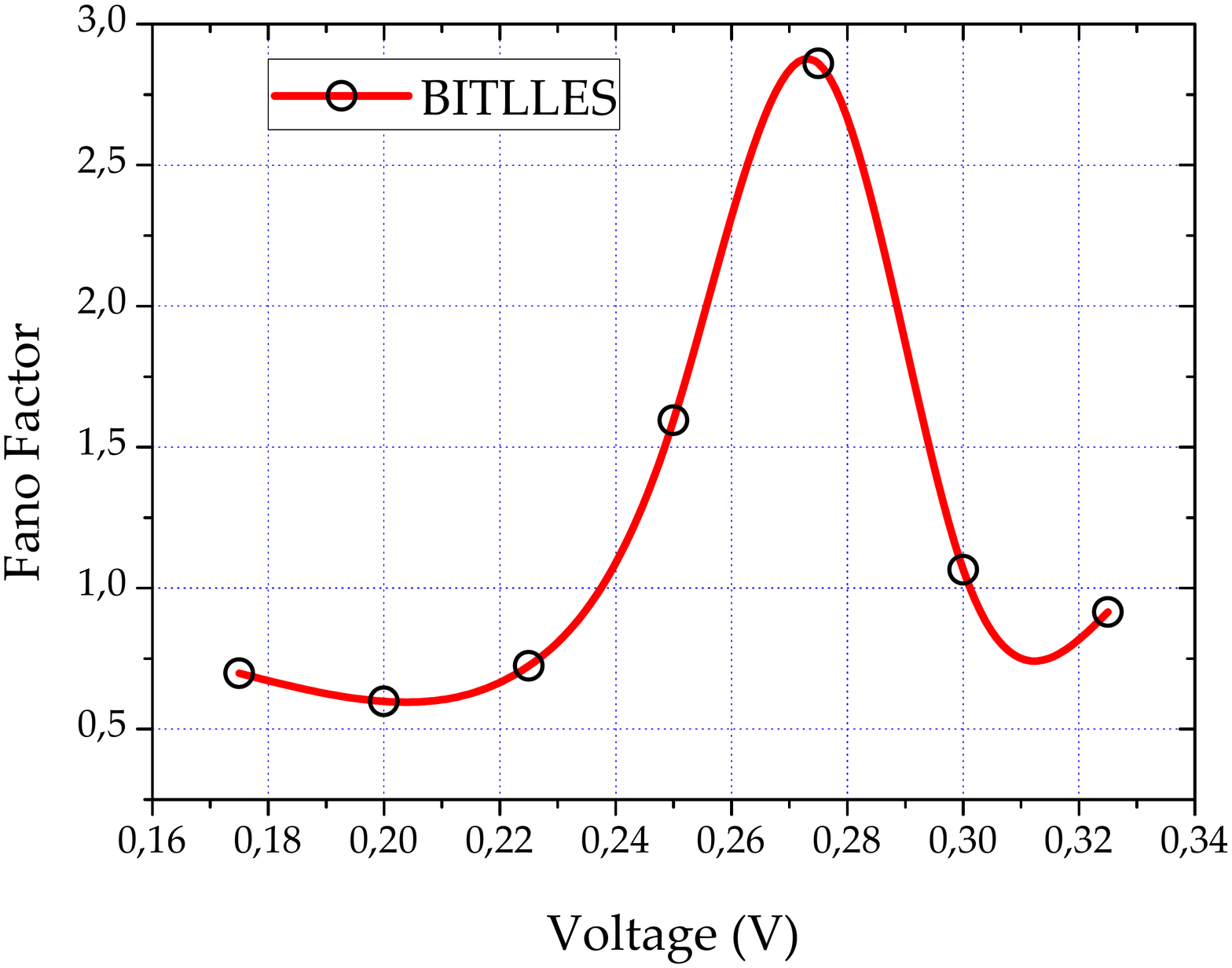}
\caption
{%
(Color online) Fano Factor F defined as $F=S(0)/(2q\left\langle I\right\rangle)$, evaluated using the current fluctuations directly available from the BITLLES simulator.}
\label{o.figure12}
\end{figure}

\begin{figure}
\centering
\includegraphics[width=0.57\columnwidth]{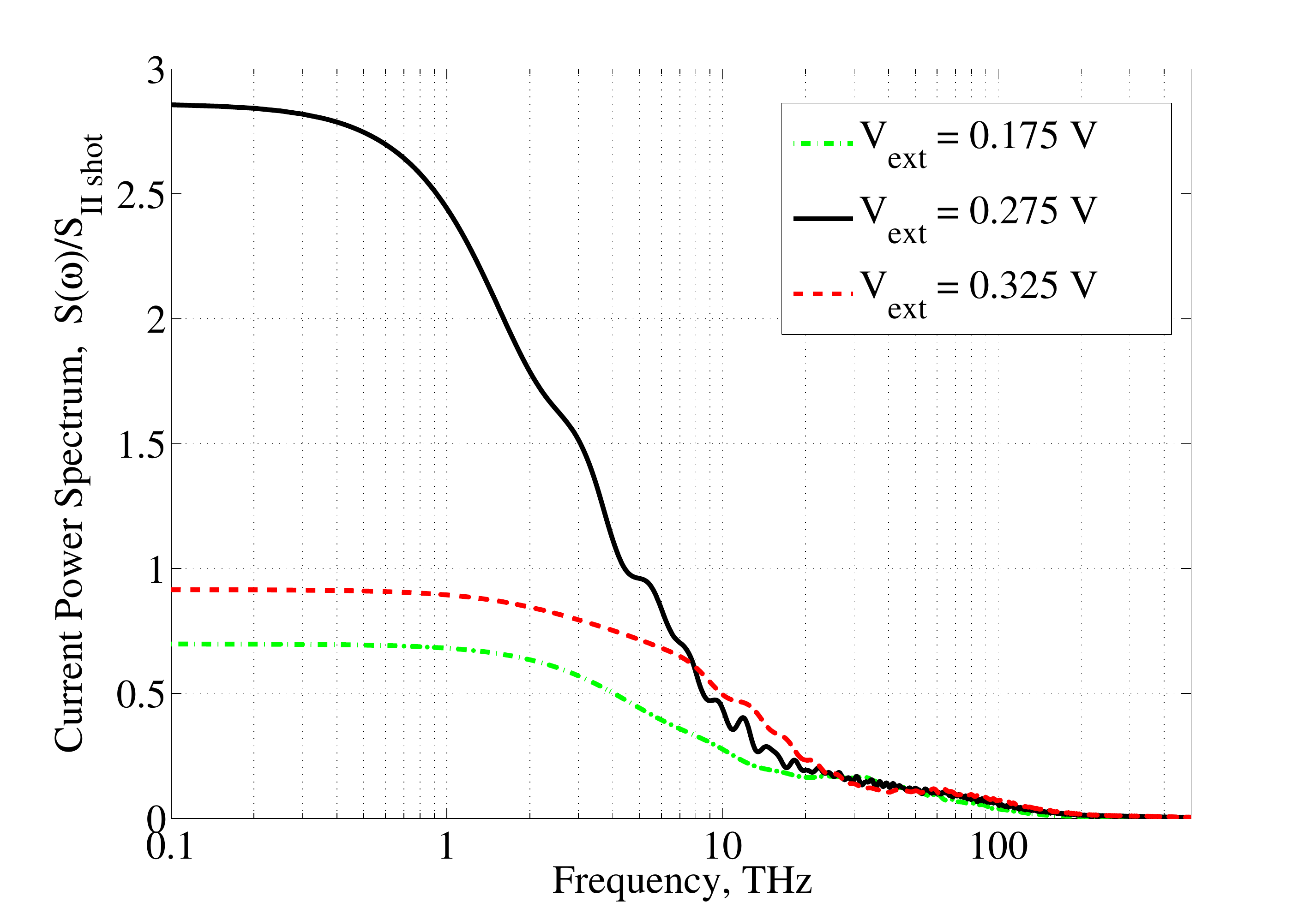}
\caption
{(Color online) Current noise power spectrum referred to Poissonian shot noise at different biases. }
\label{o.figure13}
\end{figure}

In order to understand how the many-body Coulomb interaction affects the noise in RTDs, we also investigate the correlation between an electron trapped in the resonant state during a dwell time $\tau_{d}$ and those remaining in the left reservoir.
This correlation exists essentially because the trapped electron perturbs the potential energy felt by the electrons in the reservoir. In the limit of non-interacting electrons, the Fano factor will be essentially proportional to the partition noise, however, if the dynamical Coulomb correlations are included in the simulations (see \fref{o.figure12}) this result is no longer reached, 
and hence the Fano factor is superpoissonian. Finally, we are also interested in the high frequency spectrum $S(f)$ given by (\ref{alarcon:Equation_s(f)}) revealing information about the internal energy scales of the RTD that is not available from DC transport (see \fref{o.figure13}).

Since expressions \eref{Ramo1} and \eref{Ramo2} provide information on the \emph{continuously measured} (total) electrical current at any time, any higher moment can be trivially computed from them by simply rewriting expression \eref{alarcon:Equation6_16} in terms of a series of times. It is in this regard that the trajectory-based approach decribed in this work provides fully time-resolved access to electron transport dynamics and electrical current statistics.

\section{Conclusions}
\label{conclusions}

The scientific community has an acceptable capability to predict the I-V characteristics (DC) of nanoelectronic devices. However, several difficulties are found when trying to complement these studies with AC, transient or noise behaviour. The latter predictions are equally (or even more) important than DC predictions when assessing the final role that new devices will play in the near future. The drawbacks found when trying to go beyond the DC regime are mainly related with the measurement and the many-body problems.

Firstly, the necessity to compute current-current correlations to provide information about the higher moments of the electrical current requires to deal with the effect of measurement process on the system, and it is not obvious at all how to determine the effects of the measuring process on the behaviour of the electronic system in terms of the collapse of the wave function. We have shown in Secs. \ref{Bohm2} and \ref{Bitlles1} that Bohmian mechanics provides a reasonable solution to this problem by explicitly merging waves and particles and its equations of motion. Moreover, we argued that the back action of the ammeter on the system can be sometimes disregarded without loosing much accuracy.

Secondly, the sensitivity of the electrical current to many-body correlations increases dramatically when reducing the dimensions of the electronic devices below a few nanometers and increasing their operation frequencies above the GHz regime. Physical constrictions such as overall charge neutrality or total current conservation, closely related with the carrier-carrier correlations, become crucial to make reasonable predictions of the conduction and displacement currents. The use of the so called conditional (Bohmian) wave function (see \sref{Bohm3}) and its equation of motion has been demonstrated to be a powerful strategy to deal with the many-body Coulomb correlations when combined with a proper definition of the (many-body) Poisson equation (see \sref{Bitlles2}).

Finally, the use of the above trajectory-based concepts within a quantum open system allows us to build a quantum Monte Carlo algorithm where the stochastic nature of the electrical current (\sref{Bitlles3}) is finally attributed to the number of electrons, their energies and the initial positions of the Bohmian trajectories. Based on this machinery we have developed an electron transport simulator (BITLLES) that gives access to \emph{time-resolved} electron dynamics and, in particular, to any moment of the electrical current (\sref{Computation}). Certainly, this quantum (Bohmian) trajectory approach is still in its infancy and much work is still needed. 
In particular, a proper modeling of decoherent phenomena and a description of band structure beyond effective mass are still
missing. Some preliminary works in these directions can be found in \cite{oriols_dissipation,oriols_unfalsifiable}. In any case, we have shown that Bohmian mechanics offers a powerful formalism to study quantum devices with the capabilities (DC, AC, noise, transients) that the semi-classical Monte Carlo solution of the Boltzmann equation has provided for traditional semi-classical devices.

\section{Acknowledgements}
\label{sec_Acknowledgements}
The authors acknowledge discussion with D.K. Ferry, D. Pandey, F.L. Traversa, X. Cartoix\`{a}, D. Jim\'{e}nez. 
This work has been partially supported by the \lq\lq{}Fondo Europeo de Desarrollo Regional (FEDER) and the Ministerio de Economia y
Competitividad\rq\rq{} through the Spanish Projects TEC2012-31330 and TEC2015-67462-C2-1-R, the Beatriu de Pin\'os program through the project 2010BP-A00069,
the Generalitat de Catalunya (2014 SGR-384) and by the European Union Seventh Framework
Program under the Grant Agreement no: 604391 of the Flagship initiative  \lq\lq{}Graphene-Based Revolutions in ICT and Beyond\rq\rq{}.  D.M. is supported in part by INFN and acknowledges the support of COST action (MP1006) through STSM.

\section{Appendix A: Practical algorithm to compute Bohmian trajectories}
\label{appendixA}

In this appendix, we present the simplest numerical algorithm to compute the wave function and the trajectories needed for the Bohmian formalism. Let us start by the wave function solution of the following time-dependent 1D Schr\"odinger equation:

\begin{equation}
\label{om.Schrodinger1Dappendix}
i \hbar \frac{\partial \psi(x,t)} {\partial t} = -\frac {\hbar^2} {2m^{*}} \frac{ {\partial}^2 \psi(x,t)} {\partial x^2} + U(x,t) \psi(x,t),
\end{equation}
where $\psi(x,t)$ can be understood as a single-particle wave function or the conditional wave function discussed in \eref{Conditional}. Identically, this Eq. \eref{om.Schrodinger1Dappendix} can be understood as a single-particle version of \eref{Scho} or alternatively as Eq. \eref{setpseudo}.

The first step to numerically solve the wave function is to define a mesh in time $t_j = j \Delta t$ and space $x_k = k \Delta x $ variables. Then, we define $\psi_{j}(x_k) = \psi(x,t)|_{x = x_k;t = t_j}$. The second step is to provide a finite-difference approximation for the temporal and spatial derivatives present in \eref{om.Schrodinger1Dappendix}. In particular, we use:
\begin{eqnarray}
\left.\frac{\partial \psi \left(x,t \right)}{\partial t}\right|_{x = x_k;t = t_j} &\approx& \frac{\psi_{j + 1}(x_k) - \psi_{j - 1}(x_k)} {2\Delta t}, \label{om.finite-difference_t} \\
\left.\frac{{{\partial }^{2}}\psi \left( x,t \right)}{\partial {{x}^{2}}}\right|_{x = x_k;t = t_j} &\approx& \frac{\psi_{j}(x_{k + 1}) - 2\psi_{j}(x_{k}) + \psi_{j}(x_{k - 1})}{{\Delta x}^{2}}. \label{om.finite-difference_x}
\end{eqnarray}
Inserting Eqs. \eref{om.finite-difference_t} and \eref{om.finite-difference_x} into \eref{om.Schrodinger1Dappendix}, we obtain the following simple recursive expression:
\begin{eqnarray}
\psi_{j + 1}(x_{k})=\psi_{j - 1}(x_{k}) + i\frac{\hbar \Delta t}{{{\Delta x}^{2}}m^*}\left(\psi_{j}(x_{k + 1}) - 2\psi_{j}(x_{k}) + \psi_{j}(x_{k - 1})\right)\nonumber\\
-\,i\frac{2\Delta t}{\hbar }V_{j}(x_k) \psi_{j}(x_{k}).\;\;\;\;\;\;\;\;\;
\label{om.finite-difference_recurs}
\end{eqnarray}
Once we know the wave function at the particular times $t_j$ and $t_{j - 1}$ for all spatial positions in the mesh, we can compute
the wave function for all positions at next time $t_{j + 1}$, using \eref{om.finite-difference_recurs}. The recursive application of \eref{om.finite-difference_recurs} provides the entire time evolution of the wave packet. However, we have to clarify that \eref{om.finite-difference_recurs} is not valid for the first, $x_1$, and last, $x_N$, points. To avoid discussions about the boundary conditions, we can use a very large spatial simulation box so that the entire wave packet is contained in it at any time. Then the wave function at the borders is negligible.

\begin{figure}[h]
\includegraphics[width=0.57\columnwidth]{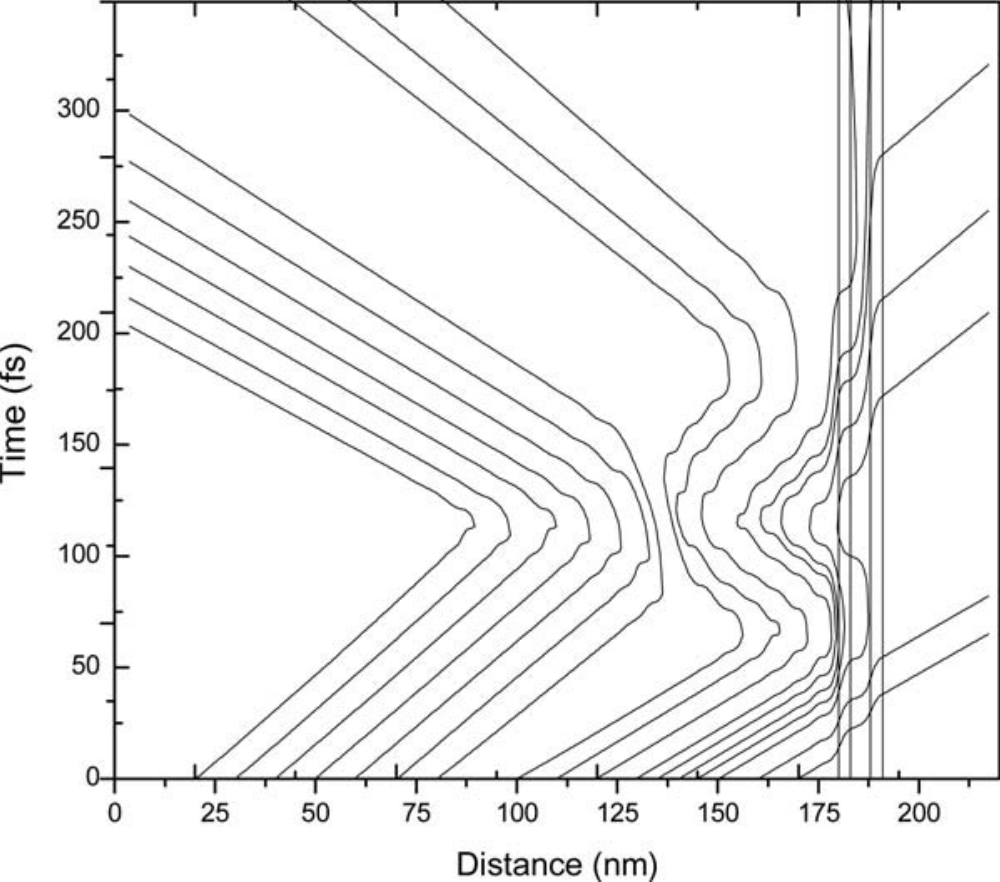}
\centering
\caption{(Color online) Representative Bohmian trajectories associated with double-packet scattering in a double-barrier potential. The position of the barriers is indicated by vertical lines.}
\label{o.figure14}
\end{figure}

\begin{figure}[h]
\includegraphics[width=0.57\columnwidth]{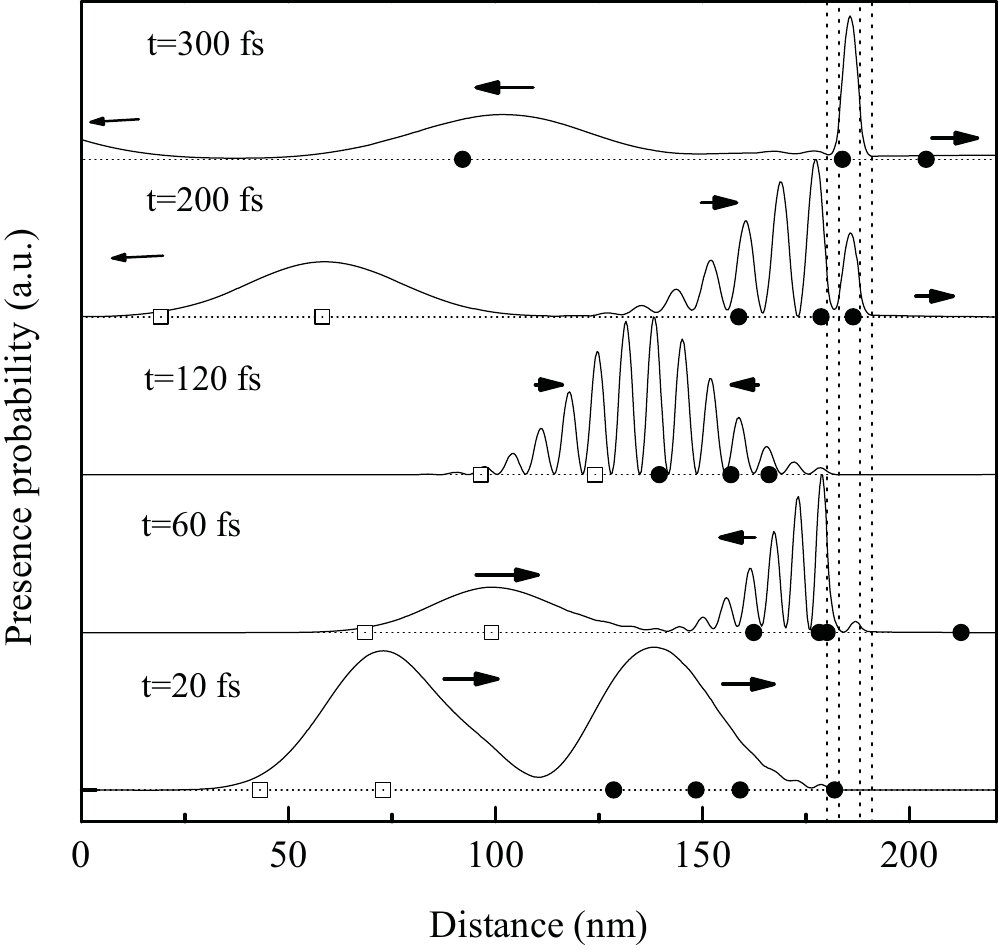}
\centering
\caption{(Color online) Picture motion of the same double-packet wave function considered in
\fref{o.figure14}, calculated by numerical integration of the
time-dependent Schr\"odinger equation. Five representative
``snapshots'' obtained at different times are shown with the
vertical scale arbitrarily changed in each case for clarity. Although
the norm of the wave function is always unity, it does not seem so
because of the changes in the scale. The marks are visual aids that
indicate the position of some related Bohmian trajectories shown in
\fref{o.figure14}. The double-barrier position is indicated by
the vertical dashed lines, and the arrows indicate the sense of
motion of the two packets.}
\label{o.figure15}
\end{figure}

This explicit solution can be unstable, and its error grows in each recursive application of \eref{om.finite-difference_recurs} \cite{o.oriols2011book}. To provide a
(conditional) stable solution we have to deal with small values of $\Delta t$ and $\Delta x$. For example, to study electron transport in nanoscale structures as depicted in Figs. \ref{o.figure14} and \ref{o.figure15}, this recursive procedure provides accurate results (the norm of the wave packets is
conserved with high precision) when $\Delta x$ is on the order of  $1-2\; \dot A$ and the temporal step, $\Delta t$, is around
$10^{-16}$s.

In order to define the initial value of the wave function, we can consider that the wave packet at the two initials times $t = \{t_1,t_1 + \Delta t\}$ evolves in a flat potential region contained in a much larger simulation box. Then, for example, we can define the  initial wave function as a time-dependent Gaussian wave packet \cite{o.cohen1978book}:
\begin{eqnarray}
\psi (x,t) ={{\left( \frac{2{{a}^{2}}}{\pi } \right)}^{1/4}}\frac{{{e}^{i\phi }}{e}^{i(k_c(x-x_0))}}{{{\left( {{a}^{4}} + \frac{4{{\hbar }^{2}}{{(t-{{t}_{1}})}^{2}}}{{{m^*}^{2}}} \right)}^{1/4}}}\times \nonumber\\
\exp \left(-\frac{{{\left[x - {{x}_0} - \frac{\hbar {{k}_{c}}}{m^*}(t - {{t}_{1}}) \right]}^{2}}}{{{a}^{2}} + \frac{2i\hbar (t - {{t}_{1}})}{m^*}} \right), \;\;\;\;\;\;\;\;\;
\label{om.finite-difference_innitial}
\end{eqnarray}
where $a$ is the spatial dispersion of the wave packet, $m^*$ the particle effective mass, $x_0$ the central position of the wave packet at the initial time $t_1$, ${k}_{c} = \sqrt{\frac{2m^*E}{{\hbar }^{2}}}$ the central wave vector in the $x$ direction related with the central energy $E$, and $\phi = -\theta - {\hbar {k_C}^{2}t}/{(2m^*)}$ with $\tan (2 \theta ) = {2 \hbar t}/{(m^*{{a}^{2}})}$ (see \cite{o.cohen1978book}). In particular, at the initial time $t = t_1$, we obtain the simplified expression:
\begin{equation}
\psi (x,t = t_1) = {{\left( \frac{2}{\pi {{a}^{2}}} \right)}^{1/4}}{{e}^{i\left( {{k}_{c}}(x-{{x}_0}) \right)}} \exp \left( -\frac{{{(x-{{x}_0})}^{2}}}{{{a}^{2}}} \right).
\end{equation}

Once we know the wave function at any position and time, we can compute the Bohmian trajectories very easily. We rewrite expression (\ref{Velocity}) to compute the velocity in a simpler form:
\begin{equation}
\label{om.velocity}
v(x,t)=\frac{J(x,t)} {|\psi(x,t)|^2}=\frac {\hbar} {m^*} Im \left( \frac {\frac {\partial \psi(x,t)} {\partial x}} {\psi(x,t)} \right),
\end{equation}
where  $J(x,t)$ is defined as:
\begin{eqnarray}
\label{om.current}
J(x,t)= i \frac {\hbar} {2m} \left( \psi(x,t) \frac {\partial \psi^{*}(x,t)} {\partial x}- \psi^{*}(x,t) \frac {\partial \psi(x,t)} {\partial x} \right)=\nonumber\\ \frac {\hbar} {m^*} Im \left( \psi^*(x,t) \frac {\partial \psi(x,t)} {\partial x} \right),
\end{eqnarray}
the square modulus $|\psi(x,t)|^2=\psi^*(x,t)\psi(x,t)$ and $Im(..)$ takes the imaginary part of a complex number. 

From the previous velocity \eref{om.velocity}, we compute each Bohmian trajectory $x(t)$ from \eref{Newtonlike}. The initial position of the Bohmian trajectory, defined here as $x(t_o)$, is selected according to the initial distribution $|\psi(x_o,t_o)|^2$. From a practical point of view, the Bohmian trajectory is not defined in a mesh neither in position nor in time. The trajectory is computed during each simulation step $\Delta t$ as:
\begin{equation}
\label{om.trajectory}
x(t_{j,k})= x(t_{{j-1},k})+v(x_k,t_{j-1}) \Delta t_k,
\end{equation}
where $\Delta t_k$ is the minimum of $\Delta t$ and the time spent by the trajectory $\{t_{j,k},t_{{j-1},k}\}$ in the cell $\{x_k,x_{k+1}\}$. If needed, we repeat the computation with a new $\Delta t_k'$ in another cell $\{x_k',x_{k'+1}\}$ and velocity  $v(x_k',t_{{j-1}.k'})$ until the total time step $\Delta t$ is finished. The Bohmian velocity in each time and position is numerically computed from the discrete wave function $\psi_j(x_{k})$ using \eref{om.velocity} as:
\begin{equation}
\label{om.velocity2}
v(x_k,t_j)=\frac {\hbar} {2\; m^* \Delta x} Im \left( \frac {\psi_j(x_{k+1})-\psi_j(x_{k-1})} {\psi_j(x_k)} \right).
\end{equation}
More complex algorithms can be found in the Appendix of chapter 1 in \cite{o.oriols2011book}. Finally, we emphasize again that the use of these trajectories extracted from \eref{om.Schrodinger1Dappendix} can seem not very useful when $\psi(x,t)$ is single-particle wave function. However,  when $\psi(x,t)$ is the conditional wave function discussed in \eref{Conditional}, then, such trajectories are extremely useful because the many-particle wave function can not be known. See the discussion in \sref{Bohm3}. 

To finish this practical explanation on how to get Bohmian trajectories from the wave function, we want to discuss some features of Bohmian trajectories related to the fact that they do not cross in the configuration space. We consider a quite ``exotic'' initial wave packet. We use a sum of two Gaussian wave packets defined by the expression \eref{om.finite-difference_innitial} with different central positions and central (momenta) wave vectors \cite{o.oriols1996pra,o.oriols2005pra}. The wave packet is certainly quite exotic because it describes just one particle. In Figs. \ref{o.figure14} and \ref{o.figure15} we see that Bohmian trajectories can be reflected for two different reasons: first, because of their interaction with the classical potential (the particles collide with the barrier) and, second, because of the collision with other trajectories traveling in the opposite direction (Bohmian trajectories do not cross). The second process is responsible for the reflection of those particles of the first packet, which never reach the barrier, and for the reflection of the entire second packet. 
These collisions between Bohmian particles are related to the quantum potential (see Ref. \cite{o.Bohm1952a}) in regions where the classical potential is zero, but for them to occur, there should be particles coming from right to left. In this regard, if the initial wave packet is prepared as a superposition of eigenstates incident from left to right (as it is always assumed in scattering thought experiments), and the classical potential is zero on the left-hand side of the barrier, then finding particles coming from the left-hand side of this region will be at least very uncommon.

\section{Appendix B: Ramo-Shockley-Pellegrini theorems}
\label{appendixB}

We describe here an algorithm to compute the total current based on the Ramo-Shockley-Pellegrini theorems \cite{o.shockley1938jap,o.ramo1939pire,o.pellegrini1986prb,o.pellegrini1993nca,o.pellegrini1993ncb}. From a practical point of view, an algorithm based on the Ramo-Shockley-Pellegrini theorems is preferred in front of \eref{displ} because it avoids some spurious numerical effects \cite{o.alarcon2009jstat} and provides an intuitive picture of the connection between the total electrical current and the geometry of the active region \cite{o.benali2012fnl,o.albareda2012fnl}.

Consider a rectangular volume $\Omega = L_x\cdot L_y\cdot L_z$ containing the whole active region drawn in \fref{o.figure4} and defined by six orthogonal surfaces $\{S_1,...,S_6\}$. A vector function $\vec{F}_{i}(\vec{r})$ inside the volume $\Omega$ ($i$ running from 1 to 6) is defined through the expression $\vec{F}_{i}(\vec{r})=-\vec{\nabla}B_{i}(\vec{r})$, where the scalar function $B_{i}(\vec{r})$ is the solution of the Laplace equation for the particular boundary condition on the surface $B_{i}(\vec{r})=1$; $\vec{r} \; \epsilon \; S_{i}$ and zero elsewhere, i.e.  $B_{i}(\vec{r})=0$; $\vec{r} \; \epsilon \; S_{h \neq i}$:
\begin{equation}
\vec{\nabla}\left( \varepsilon(\vec{r}) \vec{F}_{i}(\vec{r})\right)=-\vec{\nabla}\left( \varepsilon(\vec{r}) \vec\nabla B_{i}(\vec{r})\right)=0.
\label{alarcon:Equation6_29}
\end{equation}

The total time-dependent current through the surface $S_{i}$ can be then decomposed into two terms \cite{o.oriols2005prb}:
\begin{equation}
I_{i}(t)=\Gamma_{i}^q(t)+\Gamma_{i}^{e}(t),
\label{alarcon:Equation6_30}
\end{equation}
where:
\begin{equation}
\Gamma_{i}^{q}(t)=-\int_{\Omega}\vec{F}_{i}({\vec{r}})\vec{j}_{c}(\vec{r}_{i},t)d^{3}\vec{r},
\label{alarcon:Equation6_31}
\end{equation}
\begin{equation}
\Gamma_{i}^{e}(t)=\int_{S}\vec{F}_{i}({\vec{r}})\varepsilon(\vec{r})\frac{\partial}{\partial t}V(\vec{r},t)d\vec{s}.
\label{alarcon:Equation6_32}
\end{equation}
We use the subindex $i$ in Eqs. (\ref{alarcon:Equation6_30} - \ref{alarcon:Equation6_32}) because the current through a surface different from $S_{i}$ leads to a different definition of $\vec{F}_{i}(\vec{r})$.
Let us remark that these expressions compute the total current in a surface $S_{i}$ through a volume integral in (\ref{alarcon:Equation6_31}) and an integral over all surfaces of the volume $\Omega$ in (\ref{alarcon:Equation6_32}). The computation of $\Gamma_{i}^{q}(t)$ and $\Gamma_{i}^{e}(t)$ with trajectories can be obtained numerically from \cite{o.alarcon2009jstat}:
\begin{equation}
\Gamma_{i}^{q}(t)=\sum_{j=1}^{N(t)}\vec{F}_{i}(\vec{r}^\alpha_{j}(t))q\vec{v}_{j}(\vec{r}^\alpha_{j}(t),t),
\label{Ramo1}
\end{equation}
and
\begin{equation}
\Gamma_{i}^{e}(t)=\int_{S}\vec{F}_{i}(\vec{r})\varepsilon(\vec{r})\frac{\partial V(\vec{r},t)}{\partial t}d\vec{s}.
\label{Ramo2}
\end{equation}
Let us mention that \eref{Ramo1} not only contains the conduction current, but also contains part of the displacement current. Therefore, the numerical evaluation of the total current through a particular surface $S_i$ due to a set of $N(t)$ quantum trajectories (as discussed in \sref{Bitlles1}) can be computed from Eqs. (\ref{Ramo1}) and (\ref{Ramo2}) \cite{o.alarcon2009jstat,o.albareda2012fnl}. In fact, for a two terminal device we have to consider $L_x\ll L_y,L_z$ in order to ensure that all electric fields end finally in one of the two cables and are captured by the two surfaces of $\Omega$ close to the cables where the current is computed. Then, the shape of $B_1 (\vec r)$ in the $x$ direction is linear (we have consider $S_1=S_D$ in the \fref{o.figure4}) and a good approximation for the function $\vec F_1(\vec r)$  is:
\begin{equation}
\vec F_1(\vec r) \cdot \vec x \approx \frac {1} {L_x}.
\label{equation_linear}
\end{equation}
This is exactly the geometry discussed by Ramo and Shockley in the vacuum tube \cite{o.ramo1939pire,o.shockley1938jap}. Finally, if we assume that the surface $S_1$ and $S_4$ are in a metallic contact where there are no variations of potential and the rest of surfaces of $S$ are far from the active region, we can neglect \eref{Ramo2} and taking \eref{equation_linear} in \eref{Ramo1} we get the final result of \eref{ivelo}.

\section{Appendix C: Bohmian mechanics with operators}
\label{appendixC}

According to the final remark in \sref{Bohm2}, sometimes, the use of a Hermitian operator acting only on quantum system with the ability of providing the outcomes of the measurement process without the explicit simulation of the measuring apparatus is highly appreciated. Operators are not needed in Bohmian mechanics, but they can be very helpful mathematical tricks in practical computations. In this Appendix, we develop the expressions for commutating ensemble results from operators as an (infinite) sum of Bohmian trajectories. We consider an Hermitian operator $\hat{A}$ and its mean value $\avg{\hat{A}}_{\psi}$ in the position representation. Then, the mean value of this operator over the wave function $\psi(\vec r,t)$ is given by:
\begin{equation}
\label{om.orthodox_mean_value}
\avg{\hat{A}}_{\psi} = \int_{-\infty}^{\infty} \psi^{*}(\vec r,t) \hat{A} \left( \vec r,-i \hbar \frac {\partial} {\partial \vec r} \right) \psi(\vec r,t) d\vec r.
\end{equation}
Alternatively, the same mean value can be computed from Bohmian
mechanics by defining a spatial average of a ``local'' magnitude
$A_B(\vec r)$ weighted by $R^2(\vec r,t)$:
\begin{equation}
\label{om.Bohm_mean_value}
\avg{\hat{A}}_{\psi} = \int_{-\infty}^{\infty} R^{2}(\vec r,t) A_B(\vec r) d\vec r.
\end{equation}
In order to obtain the same value with Eqs. (\ref{om.orthodox_mean_value}) and (\ref{om.Bohm_mean_value}), we can easily identify the local mean value $A_B(\vec r)$ as:
\begin{equation}
\label{om.local_Bohm_mean_value}
A_B(\vec r) = Real \left( \left[\frac {\psi^{*}(x\vec r,t) \hat{A} \left( \vec r,-i\hbar \frac {\partial} {\partial \vec r} \right) \psi(\vec r,t)} {\psi^{*}(\vec r,t) \psi(\vec r,t)} \right]_{\psi(\vec r,t) = R(\vec r,t) e^{i \frac{S(\vec r,t)} {\hbar}}} \right).
\end{equation}
It is important to emphasize that the local Bohmian operators $A_B(\vec r)$ are not the eigenvalues of the operator $\hat{A}$. In general, the eigenvalues are not position dependent, while $A_B(\vec r)$ are. The expression $A_B(\vec r)$ is what is needed to compute the mean values of $\hat{A}$ with \eref{om.Bohm_mean_value}. 

For practical purposes, we will compute the mean value using \eref{om.Bohm_mean_value} with a large $\alpha = 1,\ldots,M_{\alpha}$ number of Bohmian trajectories with different initial positions. We will select the initial position $\vec r^\alpha(t_o)$ of the Bohmian trajectories according to the quantum equilibrium condition. Therefore, we can use \eref{Equil_hypo} to write $R^2(\vec r,t)$ in \eref{om.Bohm_mean_value}. Finally, we obtain:
\begin{equation}
\label{om.meanvalue_discrete}
\avg{\hat{A}}_{\psi}=\lim_{M_{\alpha}\rightarrow\infty} \frac {1} {M_{\alpha}} \sum_{\alpha=1}^{M_{\alpha}} A_B(\vec r^\alpha(t)).
\end{equation}
By construction, in the limit $M_{\alpha}\rightarrow\infty$, the value of \eref{om.meanvalue_discrete} is identical to the value of  \eref{om.Bohm_mean_value}. Now, we provide a few examples of how some common mean values are computed from the orthodox quantum formalism and from Bohmian trajectories. First, we compute the mean value of the position:
\begin{equation}
\label{om.position_mean_0}
\avg{x}_{\psi} = \int_{-\infty}^{\infty} \psi^{*}(x,t) x \psi(x,t) dx,
\end{equation}
with $x_B(x) = x$ so that:
\begin{equation}
\label{om.position_mean_B}
\avg{x}_{\psi} = \int_{-\infty}^{\infty} R^2(x,t) x dx.
\end{equation}
Identically, the mean value of the momentum:
\begin{equation}
\label{om.momentum_mean_0}
\avg{p}_{\psi} = \int_{-\infty}^{\infty} \psi^{*}(x,t) \left(-i\hbar \frac {\partial} {\partial x}\right) \psi(x,t) dx,
\end{equation}
with $p_B(x) = {\partial S(x,t)}/{\partial x}$:
\begin{equation}
\label{om.momentum_mean_B}
\avg{p}_{\psi} = \int_{-\infty}^{\infty} R^2(x,t) \frac {\partial S(x,t)} {\partial x} dx.
\end{equation}
For the classical potential, we have:
\begin{equation}
\label{om.Potential_energy_mean_0}
\avg{V}_{\psi} = \int_{-\infty}^{\infty} \psi^{*}(x,t) V(x,t) \psi(x,t) dx,
\end{equation}
with $V_B(x) = V(x,t)$ so that:
\begin{equation}
\label{om.Potential_energy_mean_B}
\avg{V}_{\psi} = \int_{-\infty}^{\infty} R^2(x,t) V(x,t) dx.
\end{equation}
Now, we compute the mean value of the kinetic energy:
\begin{equation}
\label{om.kinetic_energy_mean_0}
\avg{K}_{\psi} = \int_{-\infty}^{\infty} \psi^{*}(x,t) \left(-\frac {\hbar^2} {2m} \frac {\partial^2} {\partial x^2}\right) \psi(x,t) dx.
\end{equation}
It is important to notice that the local mean value of the kinetic energy takes into account the Bohmian kinetic energy plus the quantum potential. In particular, $K_B(x)$ can be obtained from the expression:
\begin{equation}
\label{om.local_kinetic_energy_mean_B}
K_B(x) = Real \left( -\frac {R(x,t) e^{-iS(x,t)/\hbar} \frac{\hbar^2} {2m} \left( \frac {\partial} {\partial x} \right)^2 R(x,t) e^{iS(x,t)/\hbar}} {R^2(x,t)} \right).
\end{equation}
The real part\footnote{It can be demonstrated quite easily that the imaginary part of \eref{om.local_kinetic_energy_mean_B} is equal to the spatial derivative of the current that becomes zero when integrated over all space. We use $J(x = \pm\infty,t) = 0$, which is always valid for wave functions that are normalized to unity, but it is not true for other types of wave functions such as plane waves.} of $K_B$ is:
\begin{equation}
\label{om.local_kinetic_energy_mean_B_bis}
K_B = \frac {1} {2m}\left(\frac {\partial S(x,t)} {\partial x} \right)^2 + Q(x,t),
\end{equation}
where $Q(x,t) = -\frac{\hbar}{2m}\frac{\nabla^2 |R(x,t)|}{|R(x,t)|}$. Finally, we obtain the Bohmian expression of the mean kinetic energy of the ensemble of trajectories:
\begin{equation}
\label{om.kinetic_energy_mean_B}
\avg{K}_{\psi} = \int_{-\infty}^{\infty} R^2(x,t) \left( \frac {1} {2m} \left(\frac {\partial S(x,t)} {\partial x} \right)^2 + Q(x,t)\right) dx.
\end{equation}
In particular, if we want to compute the ensemble (Bohmian) kinetic energy (without the quantum potential $\langle Q \rangle$), using \eref{om.meanvalue_discrete}, we get:
\begin{equation}
\langle K_{B} \rangle-\langle Q \rangle =  \mathop {\lim }\limits_{M_{\alpha} \to \infty } \frac{1}{M_{\alpha}}\sum_{\alpha = 1}^{M_{\alpha}} \frac{1}{2}m^*v^2(x^\alpha(t),t).
\label{kineticbohm}
\end{equation}
Finally, we emphasize that this way of computing ensemble values can be very useful because we avoid the inclusion of the apparatus degree of freedom as we have done in \sref{Bohm2}. However, it is not always possible to know that we have selected a \emph{good} operator that perfectly describes the measuring apparatus as we have done in \sref{Bohm2}. These ideas are emphasized by Zangh\`i, Goldstein, D\"{u}rr and coworkers when they refer to the ``naive realism about operators'' \cite{o.durr1997naive,o.durr2004equilibrium,o.durr2012book}.

\section{Appendix D: Relation between the Wigner distribution function and the Bohmian trajectories}
\label{appendixD}

In this Appendix we illustrate the formal relation between the Wigner distribution function and Bohmian mechanics. Obviously, since both formulations reproduce orthodox quantum mechanics, the observable results of the Wigner distribution function are identical to the ones obtained with the Bohmian trajectories. For computational purposes, the merit of the Wigner distribution function is its ability to deal with mixed states (defined as a statistical ensemble of pure states) through a Wigner-Weyl transformation of the density matrix. The density matrix of a mixed state can be written as $\rho(x,x\rq{})=\sum_j c_j \psi_j(x) \psi^*_j(x\rq{})$ where $c_j$  specifies the fraction of the ensemble in the pure state $\psi_j(x)$. For the sake of simplicity we avoid the explicit time dependence of the wave function and $c_j$. The need for dealing with mixed states comes from the fact that we are dealing with an open system (i.e. the active region) with a partial knowledge of the many-particle wave function in the whole (active region, cables, battery, etc.) quantum system \cite{o.Frensley}.  In the Bohmian language used in this chapter, the pure state $\psi_j(x)$ is just the conditional wave function discussed in \sref{CWF} and $c_j$ is just the $h$ distribution defined in \sref{Bitlles3}. 

The Wigner distribution function for a mixed state can be defined as a Wigner-Weyl transformation of the density matrix:
\begin{eqnarray}
\label{wigner-psi}
W(x,p) = \frac{1}{h} \sum_j c_j  \int_{-\infty}^{\infty} e^{-i p y / \hbar} \psi_j(x+\frac{y}{2}) \psi_j^{*}(x-\frac{y}{2}) dy,
\end{eqnarray}
where $p$ is the classical momentum. If we write the wave function of each pure state in polar form, i.e. $\psi_j(x) = R_j(x) e^{i S_j(x)/\hbar}$, Eq. \eref{wigner-psi} becomes: 
\begin{eqnarray}
\label{wigner-RS}
W(x,p) = \nonumber\\
\frac{1}{h} \sum_j c_j \int_{-\infty}^{\infty} e^{-i p y / \hbar} R_j(x+\frac{y}{2}) e^{i S_j(x+\frac{y}{2})/\hbar} R_j(x-\frac{y}{2}) e^{-i S_j(x-\frac{y}{2})/\hbar} dy. \;\;\;\;
\end{eqnarray}
which provides a phase-space description of the quantum system. In addition to the orthodox position distribution $\rho(x) = R^{2}(x)$, the Wigner distribution function provides a position-momentum distribution of the probabilities of the electrons with many similitudes with the classical Boltzmann distribution function used for the simulation of semi-classical devices. 

At this point a relevant question appears: \emph{What is the relation between the momentum information provided by the Wigner distribution function and the (local) velocity provided by the Bohmian theory?}. In this Appendix we want to establish this connection. We show that the average (linear) momentum at a given position $x$ computed from the Wigner distribution coincides with the Bohmian momentum at the same position, i.e. $\bar{p}(x)_W = p_B(x) = \partial S(x)/ \partial x$. \\
The space conditional momentum $\bar{p}(x)_W$ can be written in terms of the Wigner function \cite{Hiley,Molay} as:
\begin{eqnarray}
\label{momentum}
\bar{p}_W (x)= \frac{\int_{-\infty}^{\infty} p W(x,p) dp}{\int_{-\infty}^{\infty} W(x,p) dp}=\frac{\int_{-\infty}^{\infty} p W(x,p) dp}{\rho(x)}.
\end{eqnarray}
where $\rho(x) = R^{2}(x)$. 
In order to avoid unnecessary complex notation, we just focus on one pure state. The generalization to mixed states can be done straightforwardly.  Using the chain rule for a function $F(x,y)$ derivable and zero-valued at $y \to \pm \infty$, we can use the following relation:
\begin{eqnarray}
\int_{-\infty}^{\infty} dy e^{-ipy/\hbar} \frac{\partial}{\partial y} F(x,y) = \frac{i}{\hbar}p \int_{-\infty}^{\infty} dy e^{-ipy/\hbar} F(x,y),
\end{eqnarray}
and then rewrite the numerator of the right hand side of Eq. \eref{momentum} as:
\begin{eqnarray}
\label{eq-numerator-wigner}
\int_{-\infty}^{\infty} p W(x,p) dp = \frac{-i}{2\pi} \int_{-\infty}^{\infty} dp \int_{-\infty}^{\infty} dy e^{-ipy/\hbar} \nonumber \\
\cdot \Big[ R(x+\frac{y}{2})R(x-\frac{y}{2}) e^{\frac{i}{\hbar}\left[ S(x+\frac{y}{2})-S(x-\frac{y}{2})\right]} \frac{i}{\hbar}\Big( \frac{\partial S(x+\frac{y}{2})}{\partial y} -\frac{\partial S(x-\frac{y}{2})}{\partial y} \Big) \nonumber \\
+ e^{\frac{i}{\hbar}\left[ S(x+\frac{y}{2})-S(x-\frac{y}{2})\right]} \frac{\partial}{\partial y} \Big( R(x+\frac{y}{2})R(x-\frac{y}{2}) \Big) \Big]. \;
\end{eqnarray}

To proceed, let us focus on the following term:
\begin{eqnarray}
&&\frac{\partial S(x+\frac{y}{2})}{\partial y} -\frac{\partial S(x-\frac{y}{2})}{\partial y} = \nonumber \\
&=& \lim_{t \to 0} \frac{S(x+\frac{y+t}{2})-S(x+\frac{y}{2})}{t} - \lim_{t \to 0} \frac{S(x-\frac{y+t}{2})-S(x-\frac{y}{2})}{t}  \nonumber \\
&=& \frac{1}{2} \lim_{t \to 0} \frac{S(x+\frac{t}{2} + \frac{y}{2})-S(x+\frac{y}{2})}{t/2} + \frac{1}{2} \lim_{t \to 0} \frac{S(x-\frac{y}{2})-S(x-\frac{t}{2}-\frac{y}{2})}{t/2}  \nonumber \\
&=& \frac{1}{2} \left[ \frac{\partial S(x+\frac{y}{2})}{\partial x} +\frac{\partial S(x-\frac{y}{2})}{\partial x} \right],
\end{eqnarray}
which can be used to rewrite Eq. \eref{eq-numerator-wigner} as:
\begin{eqnarray}
\label{eq-num-fin}
\int_{-\infty}^{\infty} p W(x,p) dp = \int_{-\infty}^{\infty} dy \delta(y)  \cdot \nonumber\\
\Big[ R(x+\frac{y}{2})R(x-\frac{y}{2}) e^{\frac{i}{\hbar}\left[ S(x+\frac{y}{2})-S(x-\frac{y}{2})\right]} 
\frac{1}{2}\Big( \frac{\partial S(x+\frac{y}{2})}{\partial x} +\frac{\partial S(x-\frac{y}{2})}{\partial x} \Big) \nonumber\\
-i\hbar e^{\frac{i}{\hbar}\left[ S(x+\frac{y}{2})-S(x-\frac{y}{2})\right]} \frac{\partial}{\partial y} \Big( R(x+\frac{y}{2})R(x-\frac{y}{2}) \Big) \Big] = \nonumber \\
= R^{2}(x)\frac{\partial S(x)}{\partial x}. \;\;
\end{eqnarray}
In Eq. \eref{eq-num-fin}  we have used the following property:
\begin{eqnarray}
\label{delta}
\int_{-\infty}^{\infty} dp e^{-ipy/\hbar} = 2\pi \hbar \delta(y),
\end{eqnarray}
and the fact that the second term within the integral in Eq. \eref{eq-num-fin} is zero, i.e.:
\begin{eqnarray}
\label{eq-R0}
 &&\frac{\partial}{\partial y} \Big( R(x+\frac{y}{2})R(x-\frac{y}{2}) \Big) \Big|_{y=0} = \nonumber \\
 &=& \left[ \frac{1}{2} \frac{\partial R(x+\frac{y}{2})}{\partial x} R(x-\frac{y}{2}) - \frac{1}{2} R(x+\frac{y}{2})\frac{\partial R(x-\frac{y}{2})}{\partial x} \right] \Big|_{y=0} = 0.
\end{eqnarray}
Thus, plugging the result obtained in Eq. \eref{eq-num-fin} into Eq. \eref{momentum}, we can write the position dependent averaged momentum in the Wigner formalism as: 
\begin{eqnarray}
\label{eq-final-appC}
\bar{p}_W(x) = p_B (x) = \frac{\partial S(x)}{\partial x}.
\end{eqnarray}
In other words the (average) momentum at a given position $x$ computed from the Wigner function coincides with the (local) momentum defined in Bohmian mechanics. Let us come back to the mixed states mentioned at the beginning of this Appendix. It is easy to realize that a mixed state will satisfy: 
\begin{eqnarray}
\label{eq-final2-appC}
\bar{p}_W (x) = \frac {\sum_j c_j R^2_j(x) \frac{\partial S_j(x)}{\partial x} } {\sum_j c_j R^2_j(x)}.
\end{eqnarray}
The rigth hand side of \eref{eq-final2-appC} is just the weighted sum of the Bohmian momentums at the position $x$.  

Finally, let us discuss the kind of phase-space distribution that arises by gathering all Bohmian trajectories. When dealing simultaneously with several (pure) states, there will be several Bohmian velocities 
at each position $x$ (one velocity for each pure state). Therefore, as for the Wigner distribution, we will also have a (Bohmian) phase-space. Both, Wigner and Bohmian phase-space distributions are closely 
related. 
As discussed in \eref{eq-final2-appC}, the averaged momentums are identical in each position $x$.  Let us emphasize that, by construction, the phase-space distribution arising from the Bohmian velocities 
is always non-negative. The number of Bohmian trajectories with momentum $p_B(x)$ at the position $x$ has to be positive (or zero if there are no particles). 
This desirable feature of a phase-space probability distribution is not always found for the Wigner distribution. It is in this sense that we claimed along the chapter 
that the Bohmian trajectories in the BITLLES simulator for quantum devices are closely related to the Monte Carlo trajectories obtained form the Boltzmann equation for semi-classical devices.

\bibliography{biblio_Bohm_new}

\end{document}